\documentclass[journal]{IEEEtran}

\newcommand{\subparagraph}{}
\usepackage{titlesec}
\titlespacing*{\subsection}{0pt}{0.3\baselineskip}{0.2\baselineskip}
\titlespacing*{\section}{0pt}{0.5\baselineskip}{0.2\baselineskip}
\usepackage{graphicx}
\usepackage{multicol}
\usepackage[normalem]{ulem}
\usepackage{cite}
\usepackage{subfigure}
\usepackage{amsmath, amssymb, amsthm, amsfonts, latexsym, bm, subfigure, multirow, float, enumerate}
\usepackage[table]{xcolor}
\usepackage{soul}
\usepackage{adjustbox}
\usepackage{booktabs}
\usepackage{MnSymbol}
\usepackage{enumitem}
\allowdisplaybreaks

\newcommand{\STAB}[1]{\begin{tabular}{@{}c@{}}#1\end{tabular}}

\def\b{{\boldsymbol b}}

\def\j{{\boldsymbol j}} \def\J{{\boldsymbol J}}

\def\p{{\boldsymbol p}}

\def\x{{\boldsymbol x}} \def\X{{\boldsymbol X}}

\def\bbeta{\boldsymbol\beta}
\def\bkappa{\boldsymbol\kappa}

\begin{document}

\title{Bayesian Learning of Occupancy Grids}

\author{Christopher~Robbiano,~\IEEEmembership{Member,~IEEE},
Edwin~K.P.~Chong,~\IEEEmembership{Fellow,~IEEE},
Mahmood~R.~Azimi-Sadjadi,~\IEEEmembership{Life Member,~IEEE}, 
Louis~L.~Scharf,~\IEEEmembership{Life Fellow,~IEEE},
and Ali~Pezeshki,~\IEEEmembership{Member,~IEEE}       
\thanks{Manuscript was received on November 13, 2019; Revised March 15, 2020. This work was supported by the Office of Naval Research (ONR) under contract N00014-18-1-2805.

C. Robbiano, E.K.P. Chong, M.R. Azimi-Sadjadi, and A. Pezeshki are with the Electrical and Computer Engineering department at Colorado State University, Fort Collins, CO 80524 USA (email: \{chris.robbiano, edwin.chong, azimi, ali.pezeshki\}@colostate.edu).   L.L. Scharf is with the Mathematics and Statistics departments at Colorado State University, Fort Collins, CO 80521 USA (email: scharf@colostate.edu).
}}

\markboth{}%
{Robbiano \MakeLowercase{\textit{et al.}}: Bayesian Learning of Occupancy Grids}

\maketitle

\begin{abstract}
\textit{Occupancy grids} encode for hot spots on a map that is represented by a two dimensional grid of disjoint cells.  The problem is to recursively update the probability that each cell in the grid is occupied, based on a sequence of sensor measurements from a moving platform.  In this paper, we provide a new Bayesian framework for generating these probabilities that does not assume statistical independence between the occupancy state of grid cells.  This approach is made analytically tractable through the use of binary asymmetric channel models that capture the errors associated with observing the occupancy state of a grid cell. Binary-valued measurement vectors are the thresholded output of a sensor in a radar, sonar, or other sensory system.  We compare the performance of the proposed framework to that of the classical formulation for occupancy grids.  The results show that the proposed framework identifies occupancy grids with lower false alarm and miss detection rates, and requires fewer observations of the surrounding area, to generate an accurate estimate of occupancy probabilities when compared to conventional formulations.
\end{abstract}


%
\IEEEpeerreviewmaketitle

\section{Introduction}
\IEEEPARstart{T}{he} process of generating a map of occupied cells from a set of sequential observations has many applications, e.g., in target localization and path planning.  Both of these applications can be considered part of the \textit{active perception} problem, in which a sequence of actions is chosen that maximizes the amount of information attained through those actions \cite{bajcsy1988active}. An example of an active perception problem is considered in this paper, involving an ego-vehicle in the form of an autonomous underwater vehicle (AUV) which navigates through previously unexplored areas, taking a series of sequential measurements, while simultaneously performing inference for the purpose of efficiently detecting and locating underwater targets.

One of the commonly used approaches for this active perception problem is occupancy grid estimation. This process, which involves estimating the occupancy map given a set of observations, was originally introduced by Elfes and Moravec in the mid 1980s \cite{moravec1985high}. Several subsequent papers explored alternative methods for performing sensor fusion for distributing sensor measurements over the occupancy grids \cite{moravec1989sensor}, and for combining multiple occupancy grids \cite{elfes1989using, elfes1990occupancy} from multiple independent sensors into a single grid.  These methods make the assumption that the occupancy states of the grid cells are statistically independent by modeling the problem as a Markov Random Field  \cite{elfes1989using, elfes1990occupancy, moravec1985high}.  This allows for the factorization of the joint occupancy distribution on the occupancy map into the product of occupancy distributions of individual grid cells.

Thrun \cite{thrun2003learning} provided a new occupancy grid formulation, using forward sensor models, that accounts for statistical dependence between the occupancy state of grid cells.  The method assumes that the sensor provides range measurements from within an observation cone, and that the single range measurement comes from only a single source within the cone.  This measurement is assumed to be produced by either a true positive (detection of object), a false positive (false alarm), a true negative (no object in range), or a false negative (missed detection). Each of these possible events is modeled with a distribution from the exponential family, and the process of identifying the occupancy grid becomes a most-likely-model selection process through the use of an expectation-maximization (EM) algorithm \cite{thrun2003learning}.

The recent work in \cite{clarke2012sensor,guerra2018occupancy} used real antenna radiation patterns to better inform occupancy grid estimation and hence provide more realistic maps. In \cite{bai2016information,o2012gaussian}, the authors model the dependence between grid cells using Gaussian processes.  The use of Bayesian Occupancy Filters (BOF) \cite{saval2017review} for generating occupancy grids, assuming statistical independence between the grid cells, has been studied in \cite{gindele2009bayesian,coue2006bayesian,negre2014hybrid}.

The methods for occupancy grid estimation typically make the assumption that a sensor produces a range measurement, i.e., reporting the range of the closest detected object in the sensors' field of view (FoV). However, in our active perception problem, a sensory system on a moving platform takes observations at all possible ranges within its beam length, producing a vector-valued measurement. Classical occupancy grid estimation does not deal with this situation. Additionally, the classical methods treat the occupancy state of each grid cell as being statistically independent.  In a typical occupancy grid type problem, the area of a grid cell is much smaller than the area which an object occupies, leading to objects within the interrogation field typically occupying multiple grid cells. Consequently, the occupancy state of a grid cell is inherently correlated with its neighboring cells, and thus the occupancy state of the grid cells should be jointly considered when performing occupancy grid estimation. These two departures from the classical methods motivated us to develop a new occupancy grid estimation method to circumvent these problems. 

To this end, we present a framework for solving the joint and marginal distributions of grid cell occupancy states, while accounting for the statistical dependence between the occupancy states of grid cells, given vector-valued sensor measurements.  We then present a method for solving this problem with a sensor model that exploits a network of binary asymmetric channels (BACs).  This BAC network sensor model can be used to represent any real sensor, with examples given for an ideal ranging sensor, a sonar system, and a distributed network of pressure sensors.

The main contributions of this paper are as follows.  Using BACs, we build a sensor model for the interaction between the occupancy state of each grid cell and each sensor measurement.  This sensor model provides a tractable method for considering the statistical dependence between grid cell occupancy states. The use of BACs provides a method of modeling \textit{any} physical sensor by considering its statistical performance (probability of false alarm and probability of detection), and does not rely on a presumed distribution for sensor measurements. 
We also show that the original formulation of Elfes \cite{moravec1985high,elfes1989using}, which assumed statistical independence between each grid cell occupancy state, can be considered a special case of the methods proposed here.

The remainder of this paper is organized as follows. In Section \ref{sec:def}, we first define the occupancy grid estimation process and establish notation. Section \ref{sec:probform} develops the Bayesian update equation for computing the posterior probability that a particular grid cell is occupied.  Three different methods are proposed, a general formulation and two special cases.  Experimental results for each case are presented in Section \ref{sec:exp} for: a toy problem for the purpose of comparing the three proposed methods and two using simulated sonar data.  Concluding remarks are given in Section \ref{sec:conc}.

\section{Terminologies \& Notations} \label{sec:def}
Lowercase italic $x$ is used for scalars, vectors and matrices are represented by lowercase bold italic symbols $\x$ and uppercase bold italic symbols $\X$, respectively.

The environment under consideration is assumed to be partitioned into \textit{grid cells} (or simply \textit{cells}) $c_i$ at coordinates $(x_i, y_i,z_i)$, $i=1,2,\hdots, B$.  To each grid cell, an indicator variable $b_i\in\{0,1\}$ is attached with $b_i=1$ indicating that grid cell $c_i$ is \textit{occupied} by a scatterer of radiation, and $b_i=0$ indicating that grid cell $c_i$ is \textit{empty}.  An occupied grid cell may be called a hot spot.  These indicators may be organized in any convenient manner.  Here, we  arrange them into a vector $\b=\left[b_1, b_2, \hdots, b_B\right]\in\{0,1\}^B$.  It is common to call the set of indicators, organized in any manner, a \textit{map}.  A map captures the occupancy state of the grid cells.

The problem is then to estimate, for each $\bbeta\in\{0,1\}^B$, the conditional probability that $\b=\bbeta$, given a sequence of measurement vectors $\J_S=\left[\j_1, \j_2, \hdots, \j_s, \hdots, \j_S\right]$.  The subscript $S$ indicates the current sensing time, while $1\leq s \leq S$ indicates a time index at which measurements are taken. Each $\j_s \in \{0,1\}^K$ is a binary vector (e.g., thresholded detection statistics) and $\bbeta \in \{0,1\}^B$ is a specific map within the set of all possible maps with $B$ elements.  We denote this probability as $p_{\b|\J}(\b| \J_S)$ and return our estimates as the marginal posterior probabilities $p_{b|\J}(b_r=1| \J_S)$, where $b_r$ is the $r$th indicator random variable in $\b$. Importantly, it is the set of marginal posterior probabilities $\{p_{b|\J}(b_r=1 | \J_S)\}_{r=1}^B$ that is returned by our algorithm. This set of probabilities can be organized in any order, and we choose to arrange them into a vector $\p=[p_{b|\J}(b_1=1 | \J_S)$, $p_{b|\J}(b_2=1 | \J_S), \hdots, p_{b|\J}(b_B=1 | \J_S)] \in [0,1]^B$. These marginal posterior probabilities can be produced for any 1-, 2-, or 3-dimensional grid depending on the problem. 

There are some inconsistencies in the related literature as to the precise definition of an \textit{occupancy grid} or \textit{occupancy grid map} \cite{elfes1989using, elfes1990occupancy,thrun2003learning,Konolige1997,dhiman2014modern,gindele2009bayesian,coue2006bayesian,negre2014hybrid}, and whether it is a map, as defined previously, or the set of marginal posterior probabilities.  For this reason, we will refer to $\b$ as \textit{cellular occupancies}, and the set of corresponding posterior probabilities as \textit{cellular posterior probabilities}.  It is common to illustrate, or plot, both cellular occupancies and cellular posterior probabilities to provide a visual representation for comparison. Here, we reserve the occupancy grid name for both cellular occupancies and cellular posterior probabilities when they are illustrated, or plotted, in some way. 

The following is a summary of the notation used in the development of our occupancy grid estimation framework.
\begin{itemize}
     \item $B$ : The number of grid cells in a map.
    \item $\b$ : A random vector taking values in $\{0,1\}^B$ representing cellular occupancies.
    \item $\tilde{\b} \in \{0,1\}^B$ : The virtual cellular occupancies after the occupancy state of each grid cell has been passed through a binary asymmetric channel (BAC).
    \item $\p=[p_{b|\J}(b_1=1 | \J_S), \hdots, p_{b|\J}(b_B=1 | \J_S)] \in [0,1]^B$ : The vector of cellular posterior probabilities.
    \item $\mathbb{B}$ : Set of all possible cellular occupancies $\b$. $\left| \mathbb{B} \right| = 2^B$.
    \item $\mathbb{B}(r, \beta) = \{\b | b_r=\beta\}$ : Set of all cellular occupancies with $b_r=\beta$, $\beta \in \{0,1\}$.
    \item $S$ : Current/latest time index.
    \item $s$ : Time index with $1\leq s\leq S$.
    \item $\j_s \in \{0,1\}^K$ : Binary-valued measurement vector of dimension $K$ taken from the output of a sensory system at time $s$.
    \item $\J_S=\begin{bmatrix} \j_1, \hdots, \j_S \end{bmatrix}$ : Collection of binary-valued measurement vectors up to time $S$.
\end{itemize}

At each time $s$ a measurement vector $\j_s$, comprising binary values (e.g., thresholded detection statistics) $j_{s,k}$, $k=1,\ldots,K$, is taken, where $K$ is the maximum number of samples recorded by the sensor.  
The index $k$ encodes for some position $(x_k, y_k, z_k)$ within the environment. During the collection of each measurement, the sensor is effectively in a fixed position in its trajectory path, and the sensor has a FoV of the grid cells. The sequence of these measurement vectors forms the matrix $\J_S = \begin{bmatrix} \j_1, \hdots, \j_s, \hdots, \j_S \end{bmatrix}$ with columns that are all the past and present binary measurement vectors. 
Note that here we consider a sensor in its most general form to be a probabilistic map that takes as input the object we are measuring (in our case, the cellular occupancies) and gives as output a random variable (the \textit{measurement} $\j_s$). Each sensor is characterized by the conditional distribution of the output given the input.  
Real-world sensory systems can be described with this definition as presented in Section \ref{sec:ex_sonar}.

The goal is to develop a sequential Bayesian framework for updating the posterior probability of the occupancy of each grid cell $c_i$ with associated indicator random variable $b_i$,  given the sequence of measurements in $\J_S$. That is, we wish to update the cellular posterior probabilities after each time step. 

\section{Occupancy Grid Estimation-Formulation} \label{sec:probform}
To produce the estimate of the vector of cellular posterior probabilities $\p$, most of the existing methods \cite{elfes1989using,moravec1985high, elfes1990occupancy, clarke2012sensor,gindele2009bayesian} factor the joint distribution of the cellular occupancies as $p_{\b}(\b)=\prod_i p_{b}(b_i)$.  Similarly, when conditioned on the collection of measurement vectors in $\J_S$, the joint distribution is assumed to be factored as $p_{\b|\J}(\b | \J_S) = \prod_i p_{b|\J}(b_i | \J_S)$ which implies that the elements in the cellular occupancies are conditionally independent. As pointed out before, these simplifying assumptions limit the applicability of such methods in many practical situations.  The general formulation (GF) presented in this section is intended to lift these restrictive assumptions.

Throughout the development of this formulation, we assume that the sensor position at each time is reasonably accurate, i.e., the odometry errors are small enough to ignore.  This is a valid assumption for sensory system deployed on many autonomous platforms, where a low velocity and small measurement error from the odometry sensors satisfies the ego-motion estimation for the vehicle.

\subsection{Model for Grid Cell and Measurement Interactions} \label{sec:interactions}
The sensor model presented in this section provides a way to probabilistically solve the data association problem of determining which grid cells are occupied given the measurements.  A network of binary asymmetric channels (BACs) \cite{cover2012elements} are used to model the relationship between grid cell occupancy states and measurements.  The BAC outputs, the so-called \textit{virtual occupancies}, are latent variables, and are merely used to model the measurements.

\begin{figure}
    \centering
	\subfigure[Causal chain of interaction between cell occupancies $b_r$ and measurements $j_{s,k}$.]{\includegraphics[width=.48\textwidth]{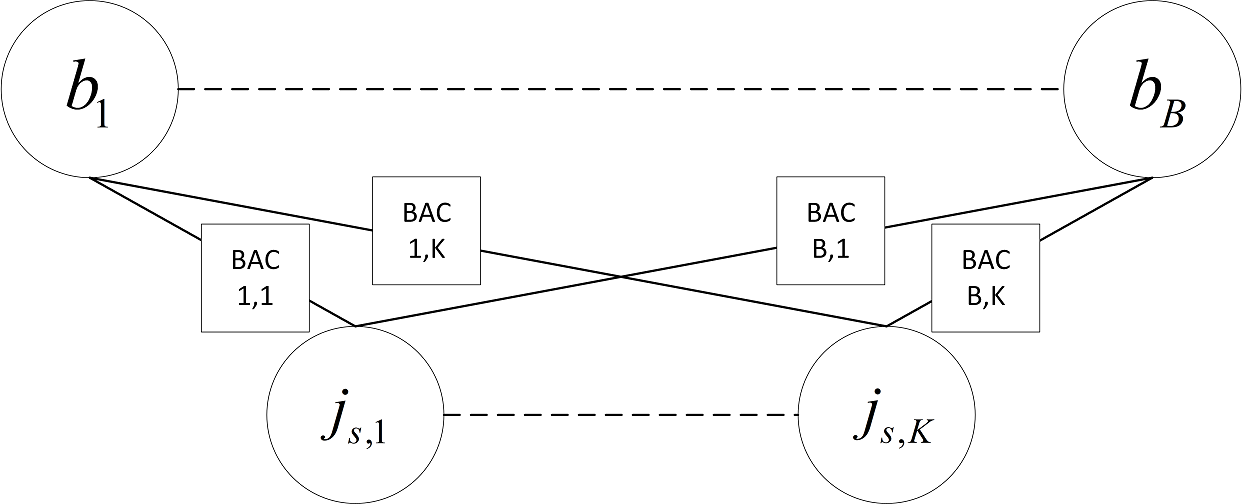}\label{fig:graphical_model}} \\
    \subfigure[Binary asymmetric channel behavior for each measurement $j_{s,k}$, conditioned on $b_r=1$.]{\includegraphics[width=.48\textwidth]{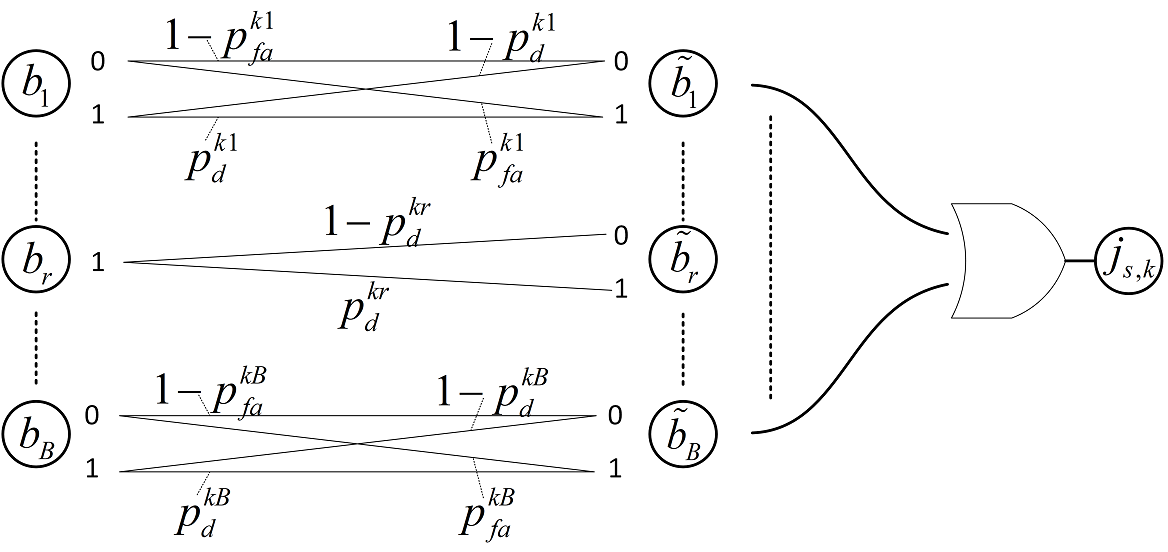} \label{fig:bac_model}}
    \caption{Interaction and measurement model}
    \label{fig:models}
\end{figure}

The occupancy state $b_i$ of each grid cell influences each $j_{s,k}$ in measurement vector $\j_s$. This relationship can be represented by a directed acyclic graph (DAG) of Figure \ref{fig:graphical_model}, where the information flow is from $b_i$ to $j_{s,k}$.  The influence that the occupancy state of each grid cell has on each measurement can be modeled by transition probabilities of a BAC. These transition probabilities can be functions of the distance between the grid cell and location of the measurement, or any other physical quantity depending on the employed sensory system, and are discussed further in Section \ref{sec:choosing}.  Recalling that $j_{s,k}$, $k=1,\hdots,K$ are binary-valued decisions from the detector at time $s$, one can view a BAC as the device that models bit flips associated with missed detection and false alarm for each grid cell.

The binary occupancy information transmitted from each $b_i$ through a BAC and received at each $j_{s,k}$ node are logical OR'd together to produce the measurement $j_{s,k}$ for all $k$, with different BAC transition probabilities for each $(i,k)$ pair. That is, letting the virtual occupancy $\tilde{b}_i$ be the output of the BAC for each grid cell $c_i$, then $j_{s,k}=\sum_{i=1}^{B} \tilde{b}_i$ where the sum is Boolean. The justification for the choice of the OR gate model is the following.  An occupied grid cell reflects a transmitted signal back to a detector, and if any of the signals received at the sensor has enough power to trigger the detector then it will be declared as detection.  This behavior is analogous to many signals being transmitted over electrical wires and connected to the circuitry that composes an OR gate.  
Figure~\ref{fig:bac_model} illustrates the graphical representation of this model in which the indicator $b_r=1$ is assigned whereas the remaining $b_i\neq b_r$ may be either a 1 or 0 before transmission through the BACs.

\subsection{Sequential Bayes' Updating} \label{sec:gf}
In this section, BAC models are used to provide a sensor model for tractably estimating the joint probability of the grid cell occupancy states \textit{without} making any of the restrictive independence assumptions stated earlier.

Given the collection of $S$ sequential observations $\J_S$,
we would like to estimate the occupancy of each grid cell. Using Bayes' rule, the update rule for the marginal probability of grid cell $c_r$ being occupied (or empty), i.e. $b_r=\beta \in \{0,1\}$, given the entire set of measurements $\J_S$, is given as
\begin{align} \label{eq:gridcellmarginal} \notag
    p_{b|\J}(b_r = 1 | \J_S) &= \frac{p_{b,\J}(b_r = 1, \J_S)}{p_{\J}(\J_S)} =  \frac{p_{b,\J}(b_r = 1, \j_S, \J_{S-1})}{p_{\J}(\j_S,\J_{S-1})} \\ \notag
    &= \frac{\left[\sum_{\b \in \mathbb{B}(r,1)} p_{\b,\J}(\b, \j_S, \J_{S-1}) \right]}{p_{\J}(\j_S,\J_{S-1})}  \\ \notag
    &=\frac{\left[\sum_{\b\in \mathbb{B}(r,1)} p_{\j|\b}(\j_S| \b,\J_{S-1}) p_{\b|\J}( \b|\J_{S-1}) \right]}{p_{\j|\J}(\j_S|\J_{S-1})} \\
    &=\eta \sum_{\b\in \mathbb{B}(r,1)} p_{\j|\b}(\j_S| \b) p_{\b|\J}( \b|\J_{S-1}),
\end{align}
where $\eta$ is a normalization term such that $p_{b|\J}(b_r=1|\J_S) + p_{b|\J}(b_r=0 | \J_S) = 1$.  The second to last line comes from the conditional independence of measurements.  The second term on the right hand side of the last line, $p_{\b|\J}(\b | \J_{S-1})$, is the posterior probability from the previous time step $S-1$.  Likewise, the posterior probability $p_{\b|\J}(\b | \J_{S})$ can be written using the same sensor model used in (\ref{eq:gridcellmarginal}):
\begin{align} \label{eq:joint_posterior}
    p_{\b|\J}(\b | \J_{S}) &= \frac{p_{\b,\J}(\j_S, \b, \J_{S-1})}{p_{\J}(\j_S, \J_{S-1})}  = \mu p_{\j|\b}(\j_S| \b) p_{\b|\J}(\b | \J_{S-1}),
\end{align}
where $\mu$ is a normalization term such that $\sum_{\bbeta \in \mathbb{B}} p_{\b|\J}(\b =\bbeta| \J_{S})=1$.  For any value of $S$, plugging (\ref{eq:joint_posterior}) into (\ref{eq:gridcellmarginal}) and expanding (\ref{eq:joint_posterior}) shows the explicit dependence on all historical measurement information for the occupancy state update.

Given the map or occupancy state $\b$, the measurements $\{j_{s,k}\}_{k=1}^K$ in  $\j_s$ are indeed conditionally independent.  Thus, we can write
\begin{align} \label{eq:jointmodel}
    p_{\j|\b}(\j_s|\b) = \prod_k p_{j|\b}(j_{s,k}|\b).
\end{align}
Note that this is different than the commonly used assumption that the elements in the cellular occupancies are conditionally independent, i.e., $p_{\b|\J}(\b | \J_S) = \prod_i p_{b|\J}(b_i | \J_S)$. 

To better describe the term $p_{j|\b}(j_{s,k}|\b)$ in (\ref{eq:jointmodel}), and specifically $p_{j|\b}(j_{s,k}=0|\b)$, let $\tilde{\b} = \{\tilde{b}_i\}_{i=1}^B$ be the collection of BAC outputs (cf. Section \ref{sec:interactions}) which are the latent variables that capture the influence of grid cell $c_i$ on each measurement $j_{s,k}$.  More specifically, each $j_{s,k}$ is a Boolean function of virtual occupancies, i.e., $j_{s,k} = \sum_{i=1}^{B} \tilde{b}_i$. Thus, we can write
\begin{align} \notag
    p_{j|\b}&(j_{s,k} =0 | \b)= p_{\tilde{b}|b}( \tilde{b}_{i} = 0 ~\forall i | \b)~:~ \text{OR gate assumption} \\  \label{eq:pjsk0}
    &= \prod_{i} p_{\tilde{b}|b}( \tilde{b}_{i} = 0 | b_{i}) = \prod_{i} p^{00}_{ki}(1-b_{i}) + p^{01}_{ki}b_{i},
\end{align}
where $p_{\tilde{b}|b}( \tilde{b}_{i} = 0 | b_{i})$ is implicitly parameterized by $k$, i.e., the index of the measurement.  For each pair $(k,i)$ of measurement index $k$ and grid cell $c_i$, we define $p_{\text{d}}^{ki}$ and $p_{\text{fa}}^{ki}$ to be the probability of detection and false alarm, respectively. Then, we have
\begin{align*}
    p^{00}_{ki}&= p_{\tilde{b}|b}(\tilde{b}_{i} = 0| b_{i} = 0) = 1-p_{\text{fa}}^{ki}, \\
    p^{01}_{ki}&= p_{\tilde{b}|b}(\tilde{b}_{i} = 0| b_{i} = 1) = 1-p_{\text{d}}^{ki},
\end{align*}
where $p_{ki}^{00}$ models the probability that $b_i=0$ is  transmitted through the BAC and received correctly as $\tilde{b}_i = 0$; while $p_{ki}^{01}$ models the probability that $b_i=0$ is transmitted through the BAC and received incorrectly as $\tilde{b}_i = 1$.  The other probabilities  $p_{ki}^{11} = p_{\text{d}}^{ki}$ and  $p_{ki}^{10} = p_{\text{fa}}^{ki}$ are similarly defined.

Plugging all these into  (\ref{eq:gridcellmarginal}) and (\ref{eq:joint_posterior}) gives the following closed-form expressions that can be used for computing the Bayesian updates:
\begin{align} \notag \label{eq:computable}
    p_{b|\J}&(b_r = 1 | \J_S) = \eta \sum_{\b\in \mathbb{B}(r,1)} p_{\j|\b}(\j_S|\b) p_{\b|\J}(\b | \J_{S-1}) \\  \notag
    &\propto \sum_{\b\in \mathbb{B}(r,1)}  \prod_k \prod_{i}\Big[ \big(p^{00}_{ki}(1-b_{i}) + p^{01}_{ki}b_{i}\big)(1-j_{S,k}) \\ 
    & \hspace{.1in} + \big(1-(p^{00}_{ki}(1-b_{i}) + p^{01}_{ki}b_{i})\big)j_{S,k} \Big] p_{\b|\J}(\b | \J_{S-1}),
\end{align}
and,
\begin{align} \label{eq:joint_computable} \notag
    p_{\b|\J}&(\b | \J_{S}) = \mu \prod_k \prod_{i}\Big[ \big(p^{00}_{ki}(1-b_{i}) + p^{01}_{ki}b_{i}\big)(1-j_{S,k}) \\
    & \hspace{.1in} + \big(1-(p^{00}_{ki}(1-b_{i}) + p^{01}_{ki}b_{i})\big)j_{S,k} \Big]  p_{\b|\J}(\b | \J_{S-1}).
\end{align}

This is what we refer to as the general formulation (GF) of the occupancy grid problem. The formulation in \cite{elfes1989using}, which assumes independence between occupancy states of grid cells, can be viewed as a special case of this formulation. To see this let us consider the case when the set $\mathbb{B}(r,1)$ contains only a single element, indicating that only one grid cell is being processed at a time.  The sum over $\mathbb{B}(r,1)$ and the product over $i$ in (\ref{eq:computable)} vanish. The product over $k$ would also be restricted to the indices $k\in \bkappa = \{\kappa, \kappa +1, \hdots, \kappa + K'\}$, with $\kappa \geq 0$ and $\kappa + K' \leq K$, that coincide with the neighborhood around grid cell $c_r$. Thus, (\ref{eq:computable}) in this case reduces to
\begin{align} \notag
    p_{b|\J}(b_r = 1 | \J_S) &= \eta \prod_{k\in \bkappa} \Big( p^{01}_{ki}(1-j_{S,k}) + \big(1-p^{01}_{ki}\big)j_{S,k} \Big) \\ \label{eq:independent}
    &\hspace{.3in} \times p_{b|\J}(b_r= 1 | \J_{S-1}),
\end{align}
which is equivalent to the formulation in \cite{elfes1989using} derived using our BAC model. 

\subsection{Choices of Transition Probabilities} \label{sec:choosing}
The choice of appropriate transition probabilities for each BAC can be made using two different categories of approaches.  The first category involves using heuristic factors where the designer imparts a belief into the choice of the transition probabilities. Such factors exploit distance-based or any other plausible measure. For example, a similar idea was adopted in the formulations presented in \cite{clarke2012sensor,guerra2018occupancy} where real-world receiver antenna gain patterns were used as a way to choose the values of $p_{ki}^{10}$ and $p_{ki}^{11}$ for different grid cells. The second category of methods uses statistical estimation methods to learn the transition probabilities given a collection of training data.  The success of these methods depends on the environments in which the senors are operating and amount of data used to estimate the transition probabilities.  For example, if the environments are consistent from experiment to experiment, and we have sufficient data, then convergence of the transition probability estimates can be  expected.  However, if the sensor takes measurements in a different environment for each experiment, then the training data may not contain sufficient information for proper estimation of the transition probabilities.

\begin{figure}
    \centering
    \includegraphics[width=.45\textwidth]{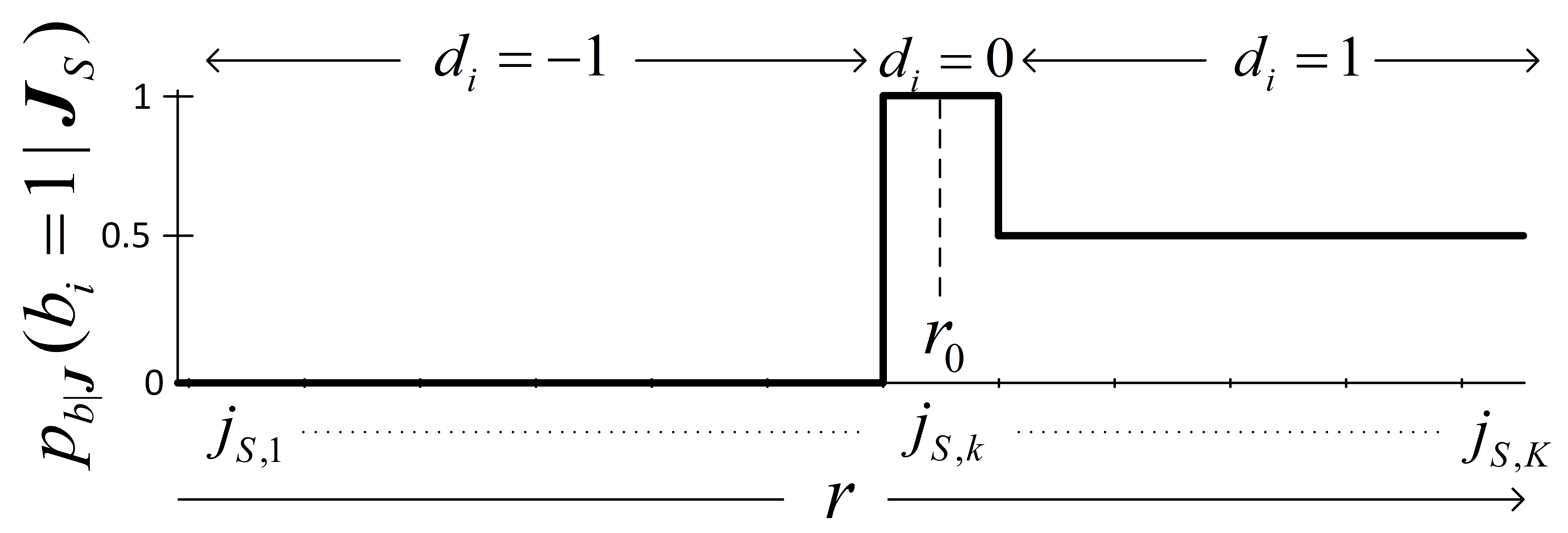}
    \caption{Occupancy probability profile for an ideal sensor.}
    \label{fig:elfes_ideal_sensor_model}
\end{figure}

We must point out that the occupancy probability profile (OPP) \cite{elfes1989using} for almost any sensor model may be represented at any time $S$ using the proposed BAC model through proper selection of the transition probabilities. As an example, we show how to model the OPP for the ideal sensor, which returns a scalar value representing the closest distance at which a reflection from an object was seen. The ideal sensor OPP \cite{elfes1989using}, illustrated in Figure \ref{fig:elfes_ideal_sensor_model}, can be modeled using the BAC model through appropriate choice of transition probabilities. This OPP modeling is performed with the use of (\ref{eq:independent}).  Let $\eta' = \eta p_{b|\J}(b_r=1|\J_{S-1})$. Then, (\ref{eq:independent}) becomes $p_{b|\J}(b_r = 1 | \J_S) = \eta' \prod_{k\in \bkappa} ( p^{01}_{ki}(1-j_{S,k}) + (1-p^{01}_{ki})j_{S,k} )$.

Let the index $k$ of $j_{S,k}$ be related to the range from the sensor, placing $j_{S,k}\in\j_S$ in ascending order by range from the sensor (i.e., $j_{S,k}$ further away from the sensor have a larger $k$). Also, let $p_{ki}^{01}$ be a function of $j_{S,k}$ and $k$. Denote by $r$ a scalar range measurement from the ideal sensor, and associate $r=r_0$ with the smallest index $k$ such that $j_{S,k}=1$.  Figure \ref{fig:elfes_ideal_sensor_model} illustrates these details along with the OPP for the ideal sensor. The scalar range measurement can be converted to its associated vector of length $K$ by assigning $0$s at all indices of the vector, then placing a $1$ at the index $k$ encoding for the distance closest to $r=r_0$.

Indicator $d_i=-1$ implies that grid cell $c_i$ with occupancy state variable $b_i$ is closer than $r_0-\varepsilon$ units from the sensor, $d_i=0$ indicate that it is $r_0\pm\varepsilon$ units from the sensor, and $d_i=1$ otherwise. The ranges associated with $d_i$ are annotated on Figure \ref{fig:elfes_ideal_sensor_model}.  The following set of rules can be applied to choose $p_{ki}^{01}$ such that $p_{b|\J}(b_i = 1 | \J_S)$ matches the OPP for the ideal sensor, where $|\bkappa|$ is the cardinality of the set $\bkappa$, and $\mathbb{I}(k)=\sum_{k\in\bkappa} 1-j_{S,k}$ is the indicator function that counts the number of $j_{S,k}\in\j_S$ that are $0$:
\begin{equation*}
    p_{ki}^{01} = 
    \begin{cases}
        0   &   d_i=-1 \\
        \frac{1 - j_{S,k}}{\eta'}^{\frac{1}{\mathbb{I}(k)}}   &   d_i=0 \\
        \frac{0.5}{\eta'}^{\frac{1}{|\bkappa|}} &   d_i=1 
    \end{cases}~ \forall k \in\bkappa.
\end{equation*}

\subsection{Special Cases for Cone-like Sensor Models} \label{sec:special_cases}
Here, we present two special cases for sensors that can be modeled with a \textit{cone-like} radiation pattern (e.g., radar, sonar, lidar, etc.).  These special cases are introduced to overcome the computational issues, specifically the exponential scaling that occurs due to the cardinality of set $\mathbb{B}(r,1)$ as the number of grid cells increases. Each special case improves computational feasibility by reducing the cardinality of $\mathbb{B}(r,1)$ such that $|\mathbb{B}(r,1)| \ll 2^{B-1}$, provided that there is a reduction in the amount of statistical dependence that is modeled between grid cells.

\begin{figure}
    \centering
    \includegraphics[width=.48\textwidth]{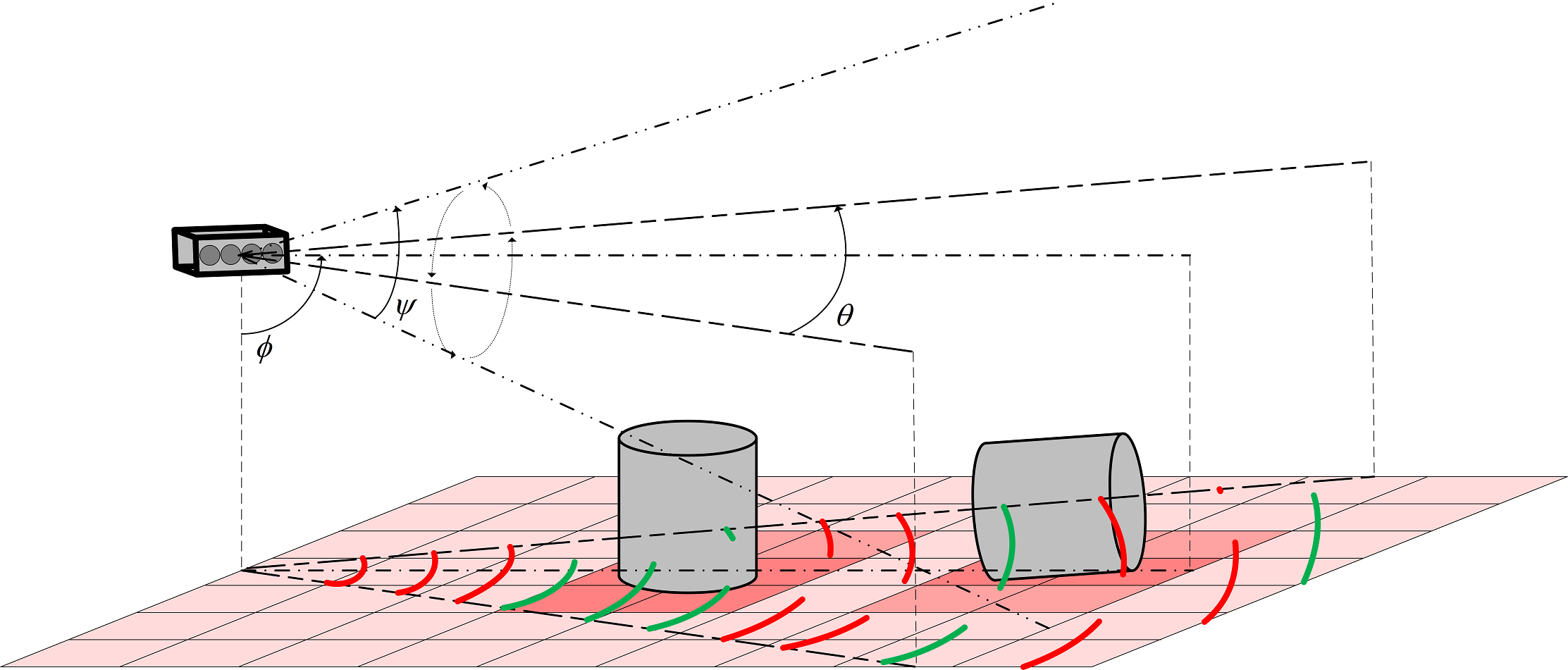}
    \caption{Sonar system and environmental setup for occupancy grid estimation at time $s$.  Uniform Linear Array (ULA) with four hydrophone elements is shown with two targets sitting on the seafloor. The sensor horizontal $\theta$ and vertical $\psi$ beamwidth, and sensor depression angle $\phi$ are shown. Colored grid cells represent the occupancy grid annotated on the seafloor, with the intensity of the color representing the probability of occupancy. Darker colors represent higher probabilities. Red dashed lines indicate $j_{s,k}=0$ and green indicate $j_{s,k}=1$.}
    \label{fig:env_setup}
\end{figure}

We show the environmental setup and application of special cases in the context of a sonar system illustrated in Figure ~\ref{fig:env_setup} as we present experiments using such a system in Section \ref{sec:ex_sonar}. Similar special cases can be constructed for other types of sensors, such as a distributed network of stationary sensors. The following physical descriptors are displayed in the figure: position of the sensor and targets at time $s$, horizontal and vertical  beamwidths $\theta$ and $\psi$, respectively, and sensor depression angle $\phi$.  Practically speaking, this information is required only for determining how to update each grid cell with each new measurement. The binary values $j_{s,k}$ are also shown, with red dashed lines indicating $j_{s,k}=0$ and green indicating $j_{s,k}=1$.

Each of the $j_{s,k}$ is generated from one of four situations: a true positive (detection of object), a false positive (false alarm), a true negative (no object in range), or a false negative (missed detection). Figure~\ref{fig:env_setup} shows these different situations in the context of a sonar system. The red- and green-dashed arcs depict the locations at which the detector has produced a zero ($j_{s,k}=0$) or a one ($j_{s,k}=1$), respectively. Thus, the presence of a green-dashed arc in an area that does not overlap with the cylindrical object in the sensor cone indicates a false alarm and the presence of a red-dashed arc in a region that does overlap with the target indicates a missed detection. An example of a false alarm can be seen in the leftmost green-dashed arc closest to the sensor in Figure~\ref{fig:env_setup}, while an example of a missed detection can be seen in the red-dashed arc intersecting the rightmost target in Figure~\ref{fig:env_setup}.

\begin{figure}
    \centering
    \subfigure[CO: Grid cell occupancy outside of the sensor cone does not influence measurements inside of the sensor cone.]{\includegraphics[width=0.23\textwidth]{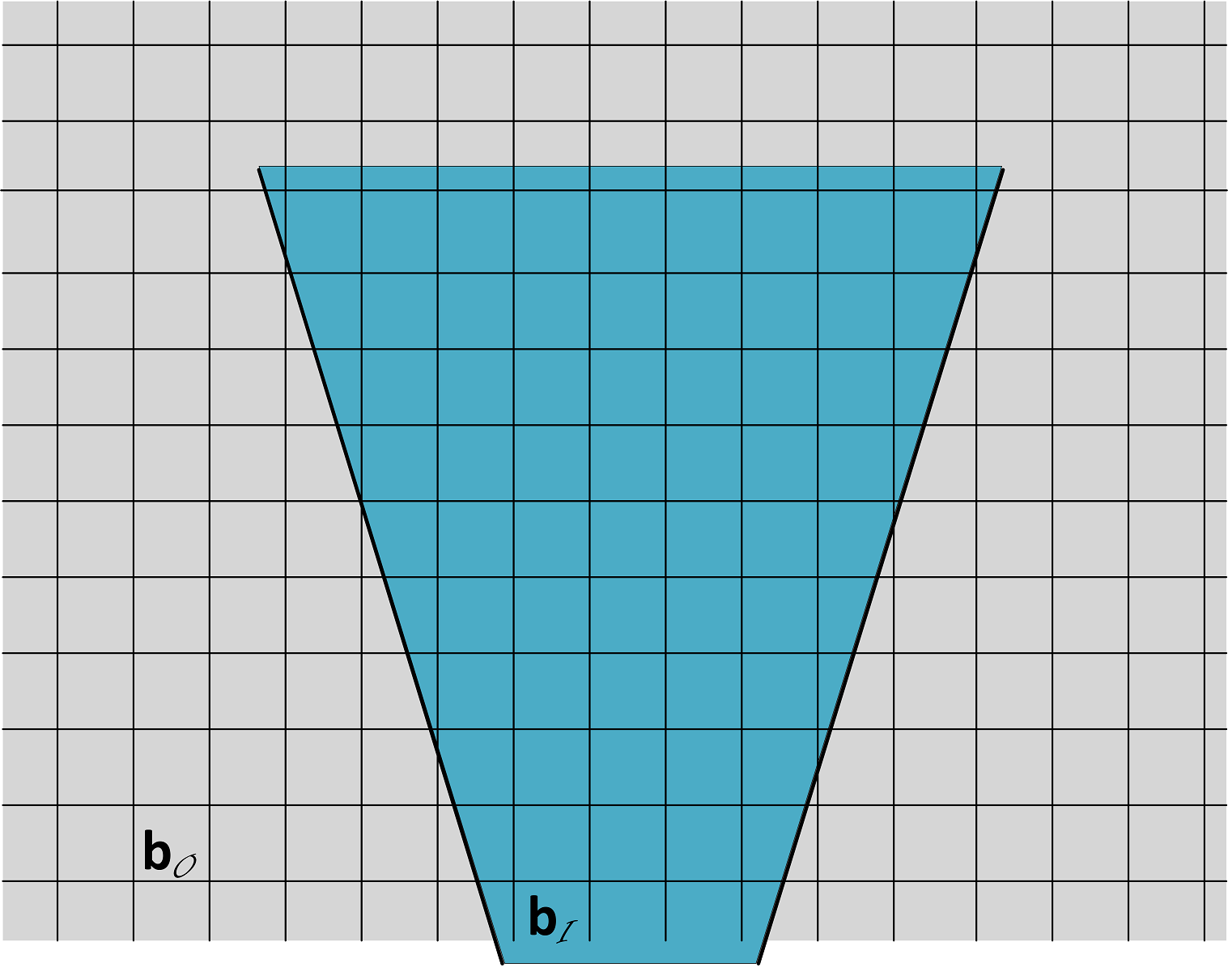} \label{fig:special_case1}} %
	\subfigure[RGO: Grid cell occupancy outside of the range gate in blue does not influence measurements inside of the range gate.]{\includegraphics[width=0.23\textwidth]{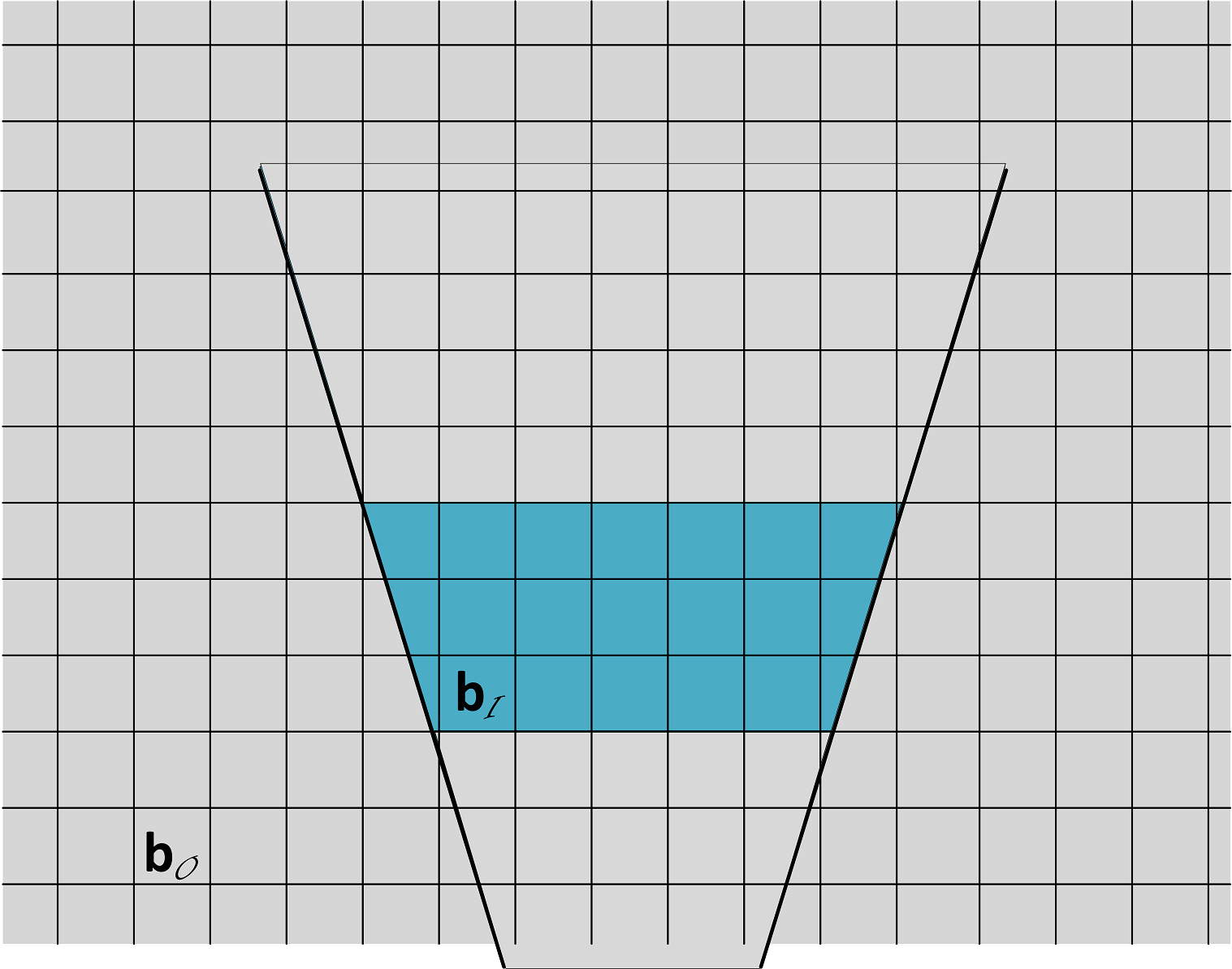} \label{fig:special_case2}}
    \caption{Physical interpretation of CO and RGO cell-measurement interactions.  The set $\b_{\mathcal{I}}$ is depicted in blue, while the set $\b_{\mathcal{O}}$ is depicted in grey for both cases.}
    \label{fig:special_cases}
\end{figure}

The first special case, which is referred to as Cone Only (CO), assumes only cone-wise interaction between the grid cells within the same observation cone at a particular time $s$ and the measurement vector $\j_s$. The second case, which is called Range Gate Only (RGO), makes the assumption that there is only local interaction between grid cells and measurements, restricted to within a particular range gate. Note that we refer to a \textit{range gate} as that comprising multiple range intervals where each \textit{range interval} is a constant distance from the sensor. A graphical interpretation of these special cases is illustrated in Figure \ref{fig:special_cases}, which helps to provide some physical intuition of each special case.

\subsection*{Case 1: Cone Only} \label{sec:special1}
Here, we assume that the interaction between the measurement vector $\j_s$ and the grid cells that lie outside the observation cone at time $s$ is negligible, and hence both the probability of detection  $p_{ki}^{11}$ and false alarm $p_{ki}^{10}$ are 0 for grid cells $c_i$ outside of the observation cone.

To formally describe this, let us partition the grid indices $\{1,2,\hdots, B\}$ into two parts, $\mathcal{I}$ and $\mathcal{O}$, such that $\mathcal{I} \bigcup \mathcal{O} = \{1,2,\hdots,B\}$ and $\mathcal{I} \bigcap \mathcal{O} = \emptyset$.  The set $\mathcal{I}$ captures the indices of grid cells that fall within the sensor cone, while the set $\mathcal{O}$ captures the indices of the grid cells that fall outside of the sensor cone. 

The indicator variables composing a particular vector $\b$ of cellular occupancies can then be partitioned into two disjoint sets, the set $\b_{\mathcal{I}}=\{b_i | i \in \mathcal{I}\}$ for grid cells \textit{inside} the sensor cone and the set $\b_{\mathcal{O}}=\{b_o | o \in \mathcal{O}\}$ for grid cells \textit{outside} such that $\b = \b_{\mathcal{I}} \bigcup \b_{\mathcal{O}}$.
Evaluating (\ref{eq:pjsk0}) for this case gives
\begin{align} \label{eq:special1} \notag
    p_{j|\b}&(j_{s,k} =0 | \b) = \prod_{o} p_{\tilde{b}|b}( \tilde{b}_{o} = 0 | b_{o})\prod_{i} p_{\tilde{b}|b}(\tilde{b}_{i} = 0 | b_{i})  \\ 
    &= (1) \prod_{i} p_{\tilde{b}|b}( \tilde{b}_{i} = 0 | b_{i}) = p_{j|\b}(j_{s,k} =0 | \b_{\mathcal{I}}),
\end{align}
because the conditional probability of receiving a $0$ through the BAC for each $\tilde{b}_{o}$ is one.  By making this assumption, the cardinality of $\mathbb{B}(r,1)$ is reduced to $|\mathbb{B}(r,1)| = 2^{B - 1 - |\mathcal{I}|}\ll 2^{B - 1}$.  Although this provides faster computational time while maintaining the statistical dependence between all grid cells within the sensor cone, it has the side effect of ignoring information imparted by neighboring occupied cells if, for example, an obstacle occupies the inside and outside of the sensor cone.

\subsection*{Case 2: Range Gate Only} \label{sec:special2}
This special case accounts for only the interaction between measurements and grid cells within a specific \textit{range gate} in the observation cone.  A visualization of a range gate can be seen in the blue band of grid cells in the observation cone in Figure \ref{fig:special_case2}.  Similar to CO, this case also ignores information imparted by neighboring grid cells if an obstacle occupies the inside and outside of a particular range gate.  A potential way to help mitigate this is to allow for overlapping range gates.

In a similar manner as was done for CO, let $\b=\b_{\mathcal{I}}\bigcup \b_{\mathcal{O}}$, but for this special case $\b_{\mathcal{I}}$ is the set of grid cells within a range gate of an observation cone as in shown in Figure \ref{fig:special_case2}.  Although (\ref{eq:special1}) can still be used for this case,  a restriction is put on $\j_{s}$ such that the delay at which the $k$th measurement is taken must coincide with the selected ranges, i.e., only the $j_{s,k}$ falling within the range gate are considered.  In this way, we calculate the posterior distributions of grid cell occupancies in each range gate separately.  This further reduces the cardinality of $\mathbb{B}(r,1)$ when compared to the first special case.

\noindent{\textit{Remarks:}}\\
The implementation of the proposed methods is expected to use the most general form that is tractable for the problem.  For a small number of total grid cells, GF can be applied without much overhead.  As the total number of grid cells increases, the cardinality of $\mathbb{B}(r,1)$ increases and thus so does the computation time for the marginalization in (\ref{eq:gridcellmarginal}).  For implementation purposes, the choice between GF, CO, and RGO should be driven by the desire to keep the cardinality of $\mathbb{B}(r,1)$ at a reasonable level that allows for a tractable amount of computation time for the system.

Consider segmenting the observation cone into many smaller disjoint cones and then using RGO to update the cellular posterior probabilities.  It is not hard to imagine that in the limit of splitting the observation cone into smaller cones and applying RGO to small range gates, the updating of grid cells would happen independently from one another.  In other words, there would only be a single grid cell being updated at any particular time.

These simplifications produce the same Bayesian update rule, presented in (\ref{eq:independent}), that one would obtain by assuming independence between the occupancy state of grid cells, and independence between measurement errors. In this case, the computational complexity would no longer scale exponentially in $B$, at the expense of possible degradation in the performance.

\section{Experimental Results} \label{sec:exp}
In this section, we present experimental results for GF defined in (\ref{eq:computable}), as well as those for CO and RGO special cases.  Two types of experiments were conducted with an increasing number of grid cells per experiment.  The first type of experiment is considered as a toy experiment, where a small number of grid cells compose the entire map. This type of experiment is necessary for comparison between GF, CO, and RGO, as there is a limit to the number of grid cells GF can be tractably applied to because of the exponential scaling of the cardinality of $\mathbb{B}(r,1)$.  The second type of experiment uses simulated sonar data, and emulates the detection and localization component of our active perception problem.

In all experiments, we assume there is an array of sensors taking observations of the map containing no moving targets.  Using the first set of experiments we show that GF provides the best overall performance, and that each special case provides performance that is related to the number of grid cells that are updated concurrently. The sensor array produces measurements that are thresholded detection statistics in the form of binary-valued vectors.  The detector's threshold was chosen to produce a desired false alarm probability $p_{\text{fa}}$ and hence a probability of detection $p_{\text{d}}$.

In all experiments, the BAC transition probabilities $p_{ki}^{00}$ and $p_{ki}^{01}$ were modeled as $p_{ki}^{00} = (1-p_{\text{d}})/(1+\text{dist}(b_i,j_{s,k}))^{\alpha}$ and $p_{ki}^{01} = (1-p_{\text{fa}})/(1+\text{dist}(b_i,j_{s,k}))^{\alpha}$, where $\text{dist}(b_i, j_{s,k})$ represents the Euclidean distance between the location of grid cell $c_i$ and that at which sample $j_{s,k}$ was taken, and $\alpha \geq 1$. This particular modeling is used to emulate degraded detection performance due to attenuation in the signal strength as a function of distance due to losses from the transmission medium, i.e., salty water for the sonar experiments.

\subsection{Metrics Used For Performance Comparison}
To evaluate the performance of various methods, two different metrics are used.  For both metrics, we consider the true cellular occupancy and the cellular posterior probabilities to be occupancy grids to facilitate direct comparison.  To make the following notation a bit more compact, we write $p_i$ as the probability mass function for the $i$th element from the cellular posterior probability occupancy grid $\p$, $\beta_i$ is the $i$th element from the ground truth occupancy grid $\bbeta$, and $p_{\beta_i}$ is the probability mass function of $\beta_i$.
\begin{enumerate}[wide, labelwidth=!, labelindent=0pt]
    \item Similarity between the cellular posterior probability occupancy grid $\p$ to that of the true occupancy grid $\bbeta$: $\rho = \langle \bbeta,\p\rangle /||\bbeta||^{2}||\p||^{2}$, where $\langle \cdot,\cdot\rangle$ is an inner product.    Clearly, $0 \leq \rho \leq 1$, and $\rho=1$ when each cellular posterior probability in $\p$ is driven to either $0$ or $1$ and $\p = \bbeta$. 
    \item Sum of the Jensen-Shannon distance (SJSD)  $D_{\text{JS}}(p_{\beta_i}||p_i)$ \cite{lin1991divergence} over all $i$ grid cells in an occupancy grid:
    \begin{align*}
        &\text{SJSD} = \frac{1}{2}\sum_{i} D_{\text{JS}}(p_{\beta_i}||p_i)  = \sum_{i}  D_{\text{KL}}(p_{\beta_i}||M_i) + D_{\text{KL}}(p_i||M_i)\\
        &= -\sum_{i}\frac{1}{2} \Big[\sum_{x\in\mathcal{X}} p_{\beta_i}(x) \log \Big( \frac{M_i(x)}{p_{\beta_i}(x)} \Big) + \sum_{x\in\mathcal{X}} p_i(x) \log \Big( \frac{M_i(x)}{p_i(x)} \Big)\Big],
    \end{align*}
    where $M_i(x)=\frac{1}{2}\times\big(p_i(x)+p_{\beta_i}(x)\big)$, and $D_{\text{KL}}(\cdot, \cdot)$ is the Kullback-Leibler (KL) divergence \cite{lin1991divergence}.  The Jensen-Shannon distance is used in favor of the KL divergence as it is symmetric, positive, and always finite. The maximum value of SJSD is $\log(2)\times B$, with smaller values indicating that $\bbeta$ and $\p$ are similar and  SJSD =0 when $\bbeta=\p$.
\end{enumerate}
\subsection{Comparison with Other Methods}
We must note that a direct comparison to the traditional occupancy grid frameworks of \cite{elfes1990occupancy,elfes1989using} is not appropriate. This is due to the fact that those methods assume that the sensor produces a single range measurement (identifies the first peak in the return above a predetermined threshold) for each observation and fit a heuristic model for sharing these range measurements over the observed grid cells at each time step.  Additionally, a direct comparison with Thrun's method \cite{thrun2003learning} that does not make the statistical independence of grid cell assumption is not possible, as the method also relies on single range measurements in its formulation. Therefore, we benchmark the performance of the proposed methods against two similar methods, both of which follow the traditional occupancy grid mapping algorithm. That is, they use a Bayesian update \cite{murphy2019introduction} for each grid cell assuming independence between the grid cells, and independence between measurements.  The first method uses our sensor model and will henceforth be referred to as the \textit{independence method} (IM).  The second method uses the conventional occupancy grid sensor model \cite{elfes1990occupancy, thrun2003learning} and is adapted to allow for binary-valued vector measurements.  We call the second method the \textit{conventional method} (CM) throughout the remained of the paper.  

The method for performing the Bayesian update on a 2-dimensional map, using IM, for experiments that use a cone shaped observation model is summarized as follows:
\begin{enumerate}
    \item Find the global position of each grid cell observed within the observation cone at time $s$.
    \item Determine the centerline of the cone, and project the position of each observed grid cell onto the centerline.
    \item Associate each measurement $j_{s,k}$ with its distance along the centerline.  Identify the set of all projected grid cells where projections equal this centerline distance.
    \item Update the probability of each grid cell $c_r$, at location $(x_r, y_r)$, being occupied given the set of measurements on that grid cell, $\{j_{s,k}\}_{k=\kappa}^{\kappa + K'}$, using (\ref{eq:independent}).
\end{enumerate}

Similarly, the Bayesian update on a 2-dimensional map using CM follows steps 1)-3) from above, and uses the sensor model proposed in \cite{elfes1990occupancy,thrun2003learning} to perform the update.  The built-in occupancy grid estimation tools available in the navigation toolbox of Matlab R2019b were used to perform the probabilistic integration for CM.

\subsection{Experiments with Toy Problem} \label{sec:toy_ex}

The first set of experiments involves a toy problem that is sufficiently small such that all three proposed methods can be applied and compared.  The toy environment used a 2-dimensional map comprising 16 grid cells with an equal distance of $0.5$ units between the centers of individual grid cells.  Each grid cell had a total of 9 different measurements sampled uniformly throughout the area it covers.  The measurement vector $\j_s$ comprises 144 measurements $\j_s = [j_{s,1},\hdots, j_{s,144}]$, $9$ samples for each of the $16$ cells, taken at equal distance covering the same overall area as the grid cells.

This experiment can be thought of as modeling multiple application scenarios. One scenario entails using a traditional ego-vehicle that is observing an area and estimating the locations of occupied grid cells using transmit-receive sensing. Alternatively, we can consider a scenario that involves using a distributed network of stationary sensors to capture simultaneous measurements.   An example is the detection of objects in a room through a distributed network of pressure sensors embedded within the floor. The latter scenario clearly shows the flexibility of our formulation for a non-traditional use of occupancy grid estimation.

\begin{figure}
    \centering
	\subfigure[True occupancy grid]{\includegraphics[width=0.15\textwidth]{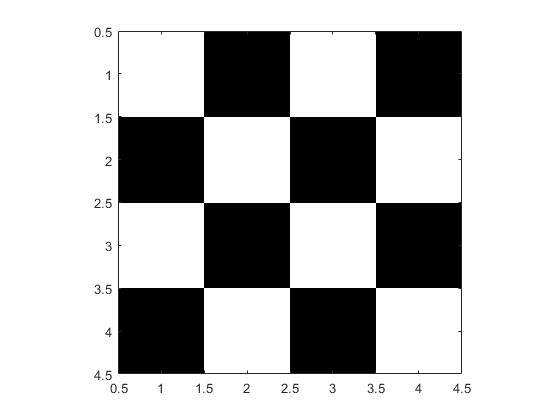} \label{fig:ex_toy1_truth}} %
	\subfigure[Example of $\j_s$ ]{\includegraphics[width=0.15\textwidth]{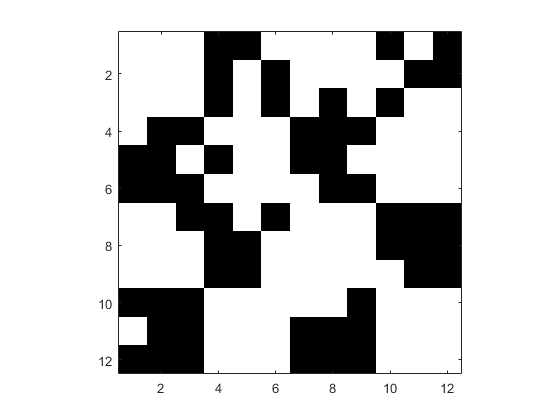} \label{fig:ex_toy1_js}}%
	\subfigure[GF occupancy grid]{\includegraphics[width=0.15\textwidth]{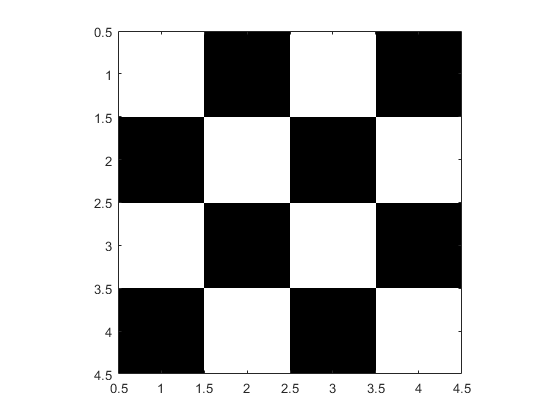} \label{fig:ex_toy1_gc}} \\
	\subfigure[CO occupancy grid]{\includegraphics[width=0.15\textwidth]{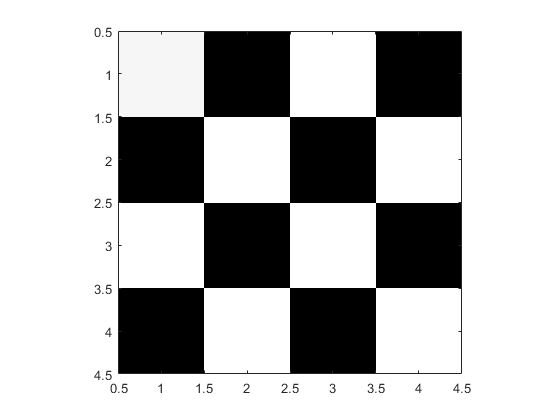} \label{fig:ex_toy1_sc1}}%
	\subfigure[RGO occupancy grid]{\includegraphics[width=0.15\textwidth]{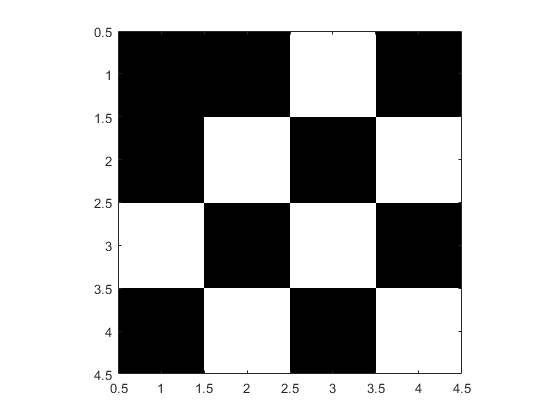} \label{fig:ex_toy1_sc2}} %
	\subfigure[IM occupancy grid]{\includegraphics[width=0.15\textwidth]{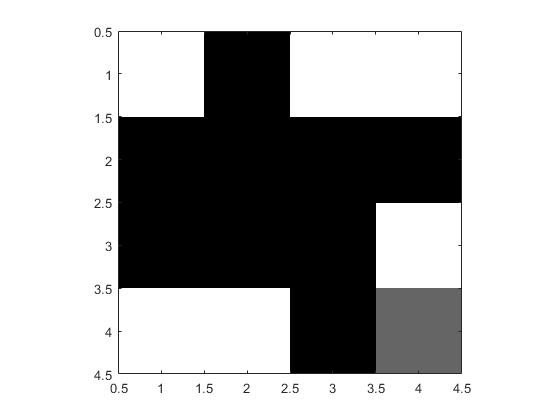} \label{fig:ex_toy1_im}}
    \caption{Checkerboard Example - True occupancy grid and cellular posterior probabilities occupancy grids generated with all three proposed methods and the independent method.}
    \label{fig:ex_toy1}
\end{figure}

The data used for the toy problem was synthesized by first choosing a $\bbeta$ (see Figure \ref{fig:ex_toy1_truth}) and finding the ideal $\j_{\text{ideal}}$ measurement vector by letting $j_{\text{ideal},k}=1$ if it corresponds to an occupied grid cell, and $j_{\text{ideal},k}=0$ otherwise.  A series of $15$ observations $\J_{15} = [\j_1, \hdots, \j_{15}]$ was then generated by passing each element of $\j_{\text{ideal}}$ through a BAC with $p_{\text{d}}=0.80$ and $p_{\text{fa}}=0.08$.  These values were chosen experimentally to match the performance from the detector used in the sonar experiments.  An example of a single measurement $\j_s$ is shown in Figure \ref{fig:ex_toy1_js}. Note that it is assumed that all grid cells are observed at all times $s$.  Each of the measurement vectors $\j_s$ was sampled randomly according to the probability law associated with the true occupancy of that grid cell.  If the grid cell ground truth is occupied, then the samples associated with that grid cell are chosen such that they are a $1$ with probability $p_{\text{d}}$ and a $0$ with probability $1-p_{\text{d}}$.  Similarly, if the grid cell ground truth is empty, then the samples associated with that grid cell are chosen such that they are a $1$ with probability $p_{\text{fa}}$ and a $0$ with probability $1-p_{\text{fa}}$.

As mentioned earlier, the BAC transition probabilities $p_{ki}^{00}$ and $p_{ki}^{01}$ were functions of the distance between the grid cell $c_i$ and the measurement location of $j_{s,k}$ with $\alpha=5$. This choice of $\alpha$ provided the best overall results for this toy problem given that there is not a physical interpretation of distance.  

Next, we compare each method of computing cellular posterior probabilities.   Special case CO is implemented on a neighborhood around grid cell locations instead of an observation cone while RGO takes the same neighborhood used for CO and splits it into two disjoint sections, updating each section separately.  Multiple experiments were conducted for each possible configuration of occupied and unoccupied grid cells ($2^{16}$ configurations), with the average performance for each method (GF, CO, RGO, and IM) reported in Table \ref{tab:toy}.  Results for CM are not presented, as they were essentially identical to those of IM for this problem.  The resulting occupancy estimates for this toy example are presented in Figures \ref{fig:ex_toy1}(c)-(f) for the GF, CO, RGO, and IM methods, respectively.   These figures illustrate the cellular posterior probabilities $\p$ where the probability of occupancy of a grid cell is represented by the gray level of the cell with a higher probability being associated with a darker cell.  

As evident from the visual evaluation of the results in Figure \ref{fig:ex_toy1} as well as the performance results in Table \ref{tab:toy}, GF outperformed all other methods in all metrics while both CO and RGO special cases outperformed the IM method. The performance of the GF and CO are essentially identical.  This is likely due to the neighborhoods of measurements used in CO being sufficiently large and including enough measurements on surrounding grid cells to accurately estimate underlying occupancy states.  RGO gave up a little performance compared to GF and CO, due to the statistical dependence that it trades off in favor of reduction in computation complexity.  It is interesting to note that the cardinality of $\mathbb{B}(r,1)$ was reduced from $2^{16}$ to $2^8$ for the CO and $2^4$ for the RGO.

\begin{table}
\centering
\caption{Experiment 1 - Comparison of cellular posterior probabilities to true occupancy grid for GF, CO, RGO, and IM.  Checkerboard (CB) and average (avg.) performance (mean$\pm$std.).}
\begin{adjustbox}{width=\columnwidth,center}
\begin{tabular}{c|c|cccc} \toprule
  & Metric & GF &CO & RGO & IM  \\
     \midrule
\multirow{2}{*}{\STAB{\rotatebox[origin=c]{90}{CB}}} & SJSD & \textbf{1.801e-3} & 1.705e-2  & 6.939e-1 & 5.1252    \\
 & $\rho$ & \textbf{1} & .9998 & .94281 & 0.57783    \\
     \midrule
\multirow{2}{*}{\STAB{\rotatebox[origin=c]{90}{Avg.}}} & SJSD & $\boldsymbol{0.58 \pm 0.51}$ & $0.60 \pm 0.51$  & $0.70 \pm 0.58$ & $2.31 \pm 1.14$    \\
    &    $\rho$ & $\boldsymbol{0.92\pm 0.07}$ & $0.92\pm 0.07$ & $0.92\pm 0.08$ & $0.70\pm 0.21$
\end{tabular}
\end{adjustbox}
\label{tab:toy}
\end{table}

\subsection{Experiments with Simulated Sonar Data} \label{sec:ex_sonar}
In our active perception problem,  a sonar system is used to search littoral zones for underwater targets e.g., mines. The experiments used a side-looking sonar (SLS) system that directs acoustic radiation to the starboard side of the AUV.  This system is equipped with multiple hydrophones arranged in a uniformly spaced linear array (ULA).  The ULA has a depression angle $\phi>0$, and horizontal $\theta$ and vertical $\psi$ beamwidths, with $\theta < \psi$, to produce a cone or wedge shaped beam as discussed in Section \ref{sec:special_cases}. The sensor cone is perpendicular to the sensor path.  For a sonar system, the width of the cone is directly related to the sonar beamwidth which is a function of the number of hydrophone array elements, aperture size, frequency, etc. The ULA has an interrogation range up to tens of meters.  The sensor array is attached to an AUV, which is some distance above the field that is under interrogation. Due to the AUV's height, the beam shape, and depression angle of the ULA, the sonar sensor tends to ``see'' past an object in the incident direction.  This is unlike some other sensor configurations, such as cameras and scanning lasers, that cannot ``see'' beyond an object in the incident direction.  This configuration allows for vector-valued measurements from each of the hydrophones in the ULA.  The AUV is assumed to be equipped with an assortment of redundant odometry sensors that provide information on AUV motion such as: heavy, surge, sway, pitch, roll, yaw, heading, depth, altitude, velocity and acceleration.  A low velocity and small measurement error from the odometry sensors satisfies the ego-motion estimation for the AUV.

The simulated sonar data used in this experiment were generated by the Personal Computer Shallow Water Acoustic Toolset (PC SWAT) simulation tool \cite{sammelmann2001propagation} developed at the Naval Surface Warfare Center, Panama City (NSWC PCD).  PC SWAT is the cutting-edge, physics-based sonar simulator that models high frequency broadband scattering from the target by a combination of the Kirchhoff approximation and the geometric theory of diffraction. Propagation of sound into a marine sediment with ripples is described by an application of Snell's law and second order perturbation theory in terms of Bragg scattering \cite{sammelmann2001propagation}.  PC SWAT has been used to produce simulations providing \textit{exemplar} template measurements that match real data generated by real shallow water sonar systems \cite{underwater2015serdp}.

The sensor data generated by PC SWAT for the ULA elements at ping $s$ were fed through an adaptive coherence estimator (ACE) detector \cite{scharf1996adaptive, kraut2001adaptive, kraut2005adaptive} to produce a single \textit{beamformed} measurement vector.  This beamformed vector is thresholded at a predetermined value to produce a desired $p_{\text{fa}}$ and $p_{\text{d}}$. The thresholded detection statistics (see Figure \ref{fig:ex_small_beta}) are then used to produce single binary-valued measurement vectors $\j_s$.

For the following experiments, the path of the AUV, and placement of the targets are illustrated in Figure \ref{fig:ex_small_truth} and \ref{fig:ex2_big_truth}.  The targets are marked by a transparent green cylindrical shape, showing the size and orientation of each target.  In some of the images, the black grid cells occlude the green rectangles (e.g., in the ground truth images).  The true occupancy grids in Figure \ref{fig:ex_small_truth} and \ref{fig:ex2_big_truth} were generated by setting $\beta_i=1$ for any grid cell that contained any significant part of a target, and a $0$ otherwise.  The path of the AUV follows the green-to-black gradient line.  The last position of the AUV is marked in a black triangle, with the AUV traveling from the dark end of the line to the triangle. After the sonar returns at each ping were received, an update to the cellular posterior probabilities takes place.

Two experiments are conducted here. The first experiment uses a short range, narrow beamwidth and coarse spatial gridding, which allows CO and RGO to be used while the second experiment uses a longer range, wider beamwidth and a finer spatial gridding, which allows only RGO to be used.  Details about each experiment will be presented in their respective sections.

\subsubsection{Experiment 1 - Short, skinny beam} \label{sec:small_ex}
In this experiment the AUV is at 10 meters above the seafloor with two cylindrical objects located 5 meters above the sandy seafloor (i.e. in the water-column) in the sonar interrogation area.  The targets were both 2 meters long along the major axis with a radius of $0.25$ meters.  The AUV uses a single sonar projector and an 11-hydrophone ULA with $3^\circ$ horizontal beamwidth.   Hydrophone elements were separated by a half-wavelength of the carrier frequency, with the geometry required to achieve the desired beamwidth being designed by PC SWAT. The transmit waveform was a linearly frequency modulated (LFM) chirp with center frequency $f_c=80$ kHz, bandwidth $BW=20$ kHz, and sampling frequency $f_s=60$ kHz.  A total of 200 pings were collected along the shown curved path, spaced at $0.01$ meters apart. 
The output for all 200 pings forms $\J_{200}=[\j_1, \hdots, \j_{200}]$.  Only CO, RGO and IM were used for this experiment owing to the very large size of $\mathbb{B}(r,1)$ for this problem.

\begin{figure}
    \centering
	\subfigure[Thresholded ACE statistics ]{\includegraphics[width=0.23\textwidth]{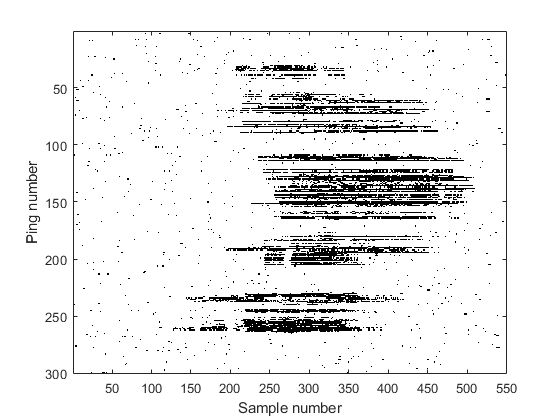} \label{fig:ex_small_beta}}%
	\subfigure[True occupancy grid]{\includegraphics[width=0.23\textwidth]{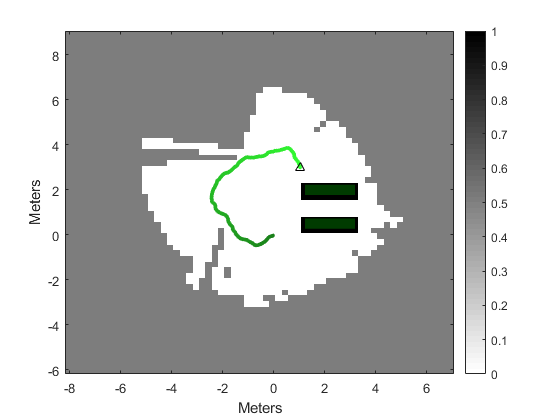} \label{fig:ex_small_truth}}
	\subfigure[CO occupancy grid]{\includegraphics[width=0.23\textwidth]{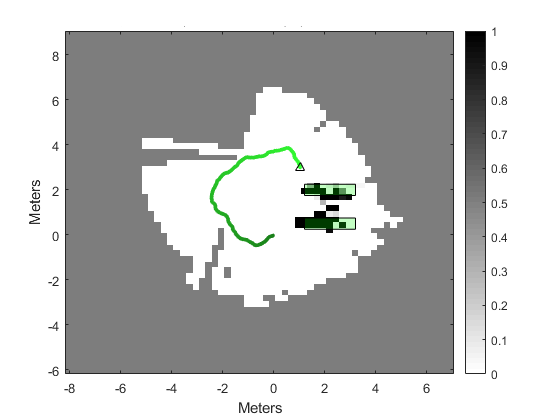} \label{fig:ex_small_sc1}}%
	\subfigure[RGO occupancy grid]{\includegraphics[width=0.23\textwidth]{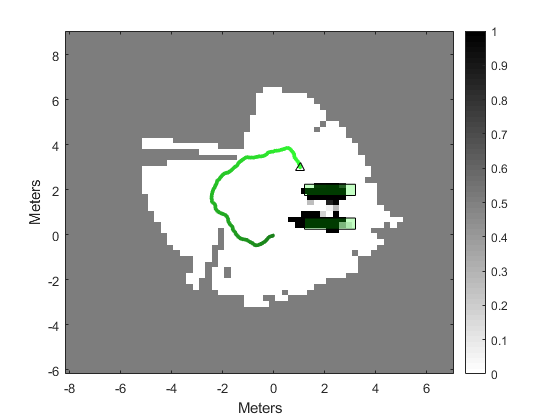} \label{fig:ex_small_sc2}}
	\subfigure[IM occupancy grid]{\includegraphics[width=0.23\textwidth]{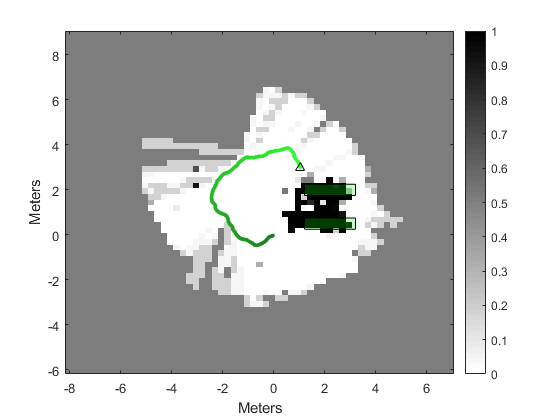} \label{fig:ex_small_im}}%
	\subfigure[CM occupancy grid]{\includegraphics[width=0.23\textwidth]{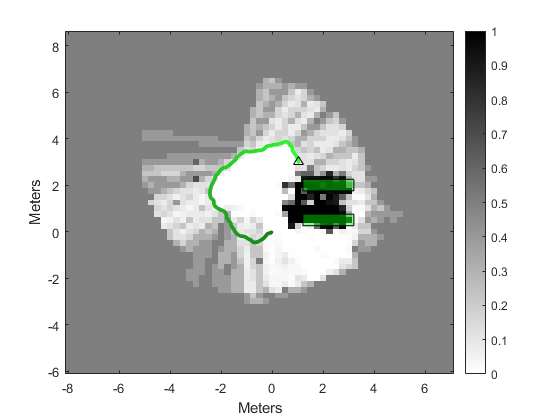} \label{fig:ex_small_cm}}

    \caption{Experiment 1 - (a) The series of measurement vectors $\j_s$, $s=1,\hdots,200$.  (b) The true occupancy grid. (c), (d), (e), (f) The occupancy grids generated by CO, RGO, IM, and CM, all with $0.25\times 0.25$ meter grid cells.  Green rectangles represent the true size, position, and rotation of the targets.  Grey-scale value of pixels represent the computed posterior occupancy probability given the measurements.}
    \label{fig:ex_small}
\end{figure}

The choice of transition probabilities plays a significant role in the performance of each method. The BAC transition probabilities $p_{ki}^{00}$ and $p_{ki}^{01}$ used for CO were chosen according to the model given earlier with $\alpha=2$.  However, for RGO these probabilities were $p_{ki}^{00}=(1-p_{\text{d}})/(1+0.96)^{2}$ and $p_{ki}^{01}=(1-p_{\text{fa}})/(1+0.96)^{2}$, i.e., did not change as a function of distance.  This constant scaling across all cells in a range gate provided better estimates for RGO.

The occupancy grids generated by the CO, RGO, and IM are illustrated in Figures \ref{fig:ex_small_sc1}, \ref{fig:ex_small_sc2}, and \ref{fig:ex_small_im}, respectively.  Six range gates were used for RGO.  We can see that both CO and RGO visually perform in a similar manner, surrounding the true occupied grid cells with areas of high probability of occupancy while providing an area of low probability of occupancy between the two targets.  IM and CM, on the other hand, could not resolve the two closely spaced targets and hence merged them together.  When compared to CO, they had more grid cells that are in an uncertain state with their posterior occupancy probabilities remaining within a small range around $0.5$, instead of being close to $0$ or $1$.
\begin{table}
\centering
\caption{Experiment 1 - Comparison of cellular posterior probabilities occupancy grids to true occupancy for CO, RGO, IM, and CM}
\begin{tabular}{c|ccccc} \toprule
 Metric & CO & RGO & IM & CM \\
     \midrule
 SJSD & 23.8306 & \textbf{20.2548} & 43.2365 & 119.99   \\
  $\rho$ & 0.9768 & \textbf{0.9805} & 0.9673 & 0.9127
\end{tabular}
\label{tab:small}
\end{table}

The performance for each of the methods was also evaluated using the SJSD and $\rho$ measures and the results are presented in Table~\ref{tab:small}. As can be seen from these results,  RGO slightly outperformed CO.  This is, in part, attributed to a smaller number of hot spots on the upper target in Figure~\ref{fig:ex_small_sc1}, which is likely due to the larger mismatch in the collection of transition probabilities for CO resulting from poorly modeling the statistical dependence between grid cell occupancy states. Both IM and CM were outperformed by CO and RGO as they produced more false alarms, even though they had a greater number of hits on each target.

\begin{figure}[h]
    \centering
	\subfigure[Experiment 1]{\includegraphics[width=0.23\textwidth]{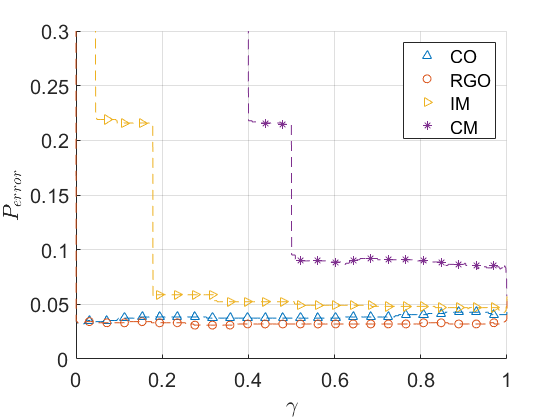} \label{fig:ex_small_pe}}%
	\subfigure[Experiment 2]{\includegraphics[width=0.23\textwidth]{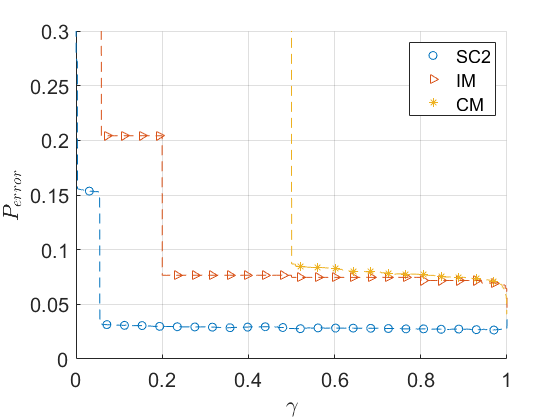} \label{fig:ex_big_pe}}

    \caption{Probability of error as a function of threshold $\gamma$ for different methods.}
    \label{fig:ex_pes}
\end{figure}

It is possible to generate an estimate of the underlying occupancy by applying a threshold $0 \leq\gamma\leq 1$ to the cellular posterior probabilities in order to produce a binary detection map that facilitates further analysis by human operator or autonomous navigation systems. An appropriate value for the threshold $\gamma$ can be chosen, e.g., to minimize the probability of error. Figure \ref{fig:ex_small_pe} shows the probability of error as a function of the threshold $\gamma$ for CO, RGO, IM, and CM.  The results presented in Table \ref{tab:small} are echoed in Figure \ref{fig:ex_small_pe}, as RGO provides a lower probability of error for \textit{all} thresholds, followed by CO, and finally IM and CM.

This result, along with the results from the previous toy experiment, suggest that using RGO provides similar performance to those of both GF and CO while outperforming the IM method. Moreover, RGO involves fewer computations than CO. Compared to IM, RGO incurs only a minor increase in computation time while providing considerably better occupancy grid estimation results. All methods outperform CM.
\subsubsection{Experiment 2 - Long, wider beam}
In many ways,  this experiment is similar to Experiment 1 in Section \ref{sec:ex_sonar}, with some exceptions discussed here. The AUV is 10 meters above the seafloor. Four cylindrical objects are partially buried and/or proud on the sandy seafloor.  The horizontal beamwidth is $10^\circ$.  A total of 300 pings were collected along a curved path, spaced at $0.1$ meters apart between pings.  The cardinality of $\mathbb{B}(r, 1)$ is greater than $2^{100}$ for an observation cone.  Thus, this precludes the use of CO, as the computation time for a single ping becomes unreasonable. Therefore, only RGO and IM were used for this experiment.  

\begin{figure}[h]
    \centering
	\subfigure[Thresholded ACE statistics]{\includegraphics[width=0.23\textwidth]{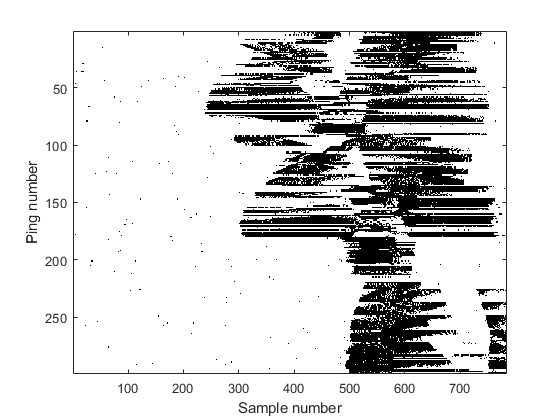} \label{fig:ex2_big_beta}}%
	\subfigure[True occupancy grid]{\includegraphics[width=0.23\textwidth]{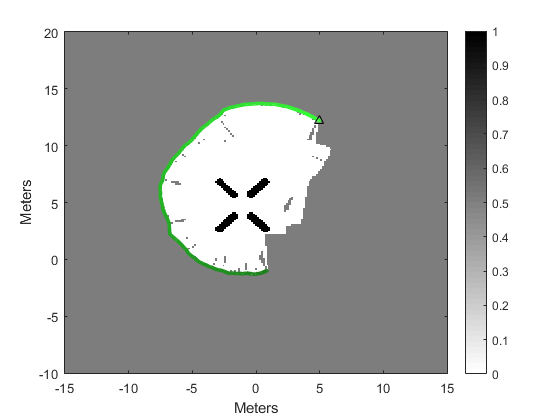} \label{fig:ex2_big_truth}}
	\subfigure[RGO occupancy grid]{\includegraphics[width=0.23\textwidth]{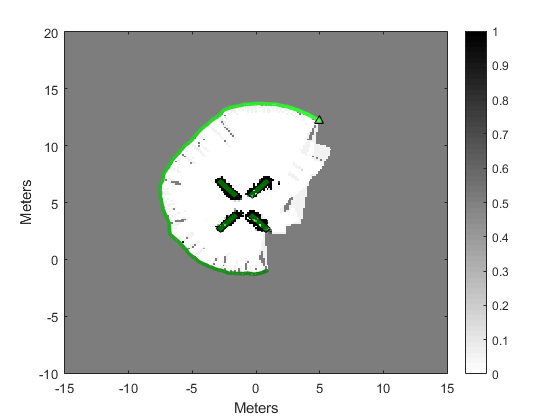} \label{fig:ex2_big_sc2}}%
	\subfigure[IM occupancy grid]{\includegraphics[width=0.23\textwidth]{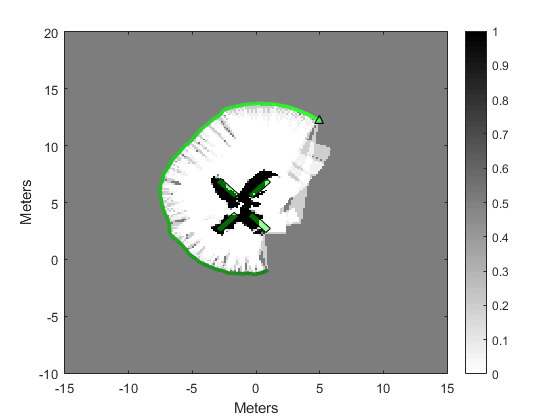} \label{fig:ex2_big_im}}
	\subfigure[CM occupancy grid]{\includegraphics[width=0.23\textwidth]{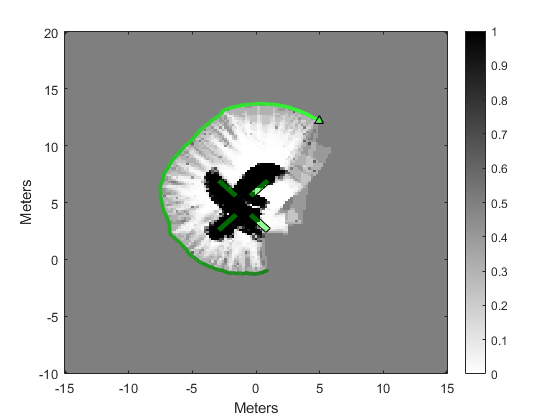} \label{fig:ex2_big_cm}}

    \caption{Experiment 2 - (a) The series of measurement vectors $\j_s$, $s=1,\hdots,200$.  (b) The true occupancy grid. (c), (d), (e) The occupancy grids generated by RGO, IM, and CM, all with $0.2\times 0.2$ meter grid cells.  Green rectangles represent the true size, position, and rotation of the targets.  Grey-scale value of pixels represent the computed posterior occupancy probability given the measurements.}
    \label{fig:ex_big}
\end{figure}

The occupancy grids generated by the RGO, and IM are illustrated in Figures \ref{fig:ex2_big_sc2} and \ref{fig:ex2_big_im}, respectively.  One hundred and thirty two overlapping range gates were used for RGO.  The BAC transition probabilities used for RGO are the same as those in Experiment 1.

From Figure~\ref{fig:ex2_big_sc2}, it is seen that RGO clearly identified the four separate targets by correctly separating them while generating hot spots over the majority of each target.  It does this without producing too many false alarms.  The IM also identifies the four targets, but produced many false alarms that join the four targets together.  It produced the correct target shapes, but offset the hot spots slightly when compared to the actual target locations. Unlike RGO and IM methods, the results of CM method shown in Figure~\ref{fig:ex2_big_cm} are unacceptable as it greatly overestimated the cellular posterior probabilities surrounding the targets. 

The values for SJSD and $\rho$ are recorded in Table~\ref{tab:big}, where RGO can be seen to outperform IM.  The value for $\rho$ is similar between the two methods, which is to be expected as both of the occupancy grids are similar to the truth.   

\begin{table}
\centering
\caption{Experiment 2 - Comparison of cellular posterior probabilities occupancy grids to true occupancy grid for RGO, IM, and CM}
\begin{tabular}{c|ccc} \toprule
 Metric & RGO & IM & CM \\
     \midrule
 SJSD &  \textbf{120.22} & 367.01 & 1019.94   \\
  $\rho$ & \textbf{0.9916} & 0.9754 & 0.9384
\end{tabular}
\label{tab:big}
\end{table}

As with Experiment 1, the probability of error as a function of the threshold $\gamma$ was computed and illustrated in Figure \ref{fig:ex_big_pe}. As can be seen, the probability of error for RGO is lower than that of IM and CM for every choice of the threshold. This indicates that the addition of some complexity to the joint distribution model in the form of inter-cell statistical dependence, albeit small in the case of RGO, gives considerable improvements over IM.

\section{Concluding Remarks} \label{sec:conc}
In this paper we have presented a new formulation for occupancy grid estimation that accounts for the statistical dependence between grid cell occupancy states and allows for vector valued measurements. This contrasts with the classical methods in \cite{elfes1989using,moravec1985high} that consider the occupancy states of grid cells to be statistically independent, as well as the work in \cite{thrun2003learning} that considers statistical dependence but only allows for scalar measurements through the use of correspondence variables.

We have shown that the independence method can be viewed as a special case of our formulation. The experimental results reveal that our formulations outperform the independence method which in turn outperforms the conventional method.  The proposed method offers much better resolution of small gaps between closely positioned objects, whereas the independent method groups them together.  Although it is intractable to use the general formulation for real-world applications, modeling some correlation effects between the occupancy state of grid cells, as offered by CO and RGO methods, indeed provides good approximation to the general form while offering significantly reduced computational time.
\section*{Acknowledgments}
The authors would like to specially thank Denton Woods and Gary Sammelmann at Naval Surface Warfare Center Panama City (NSWC PCD) for providing the  Personal Computer Shallow Water Acoustic Toolset (PC SWAT) simulator and their support throughout this research.
\bibliography{books}
\bibliographystyle{IEEEtran}
\begin{IEEEbiography}[{\includegraphics[width=1in, height=1.25in, clip,keepaspectratio]{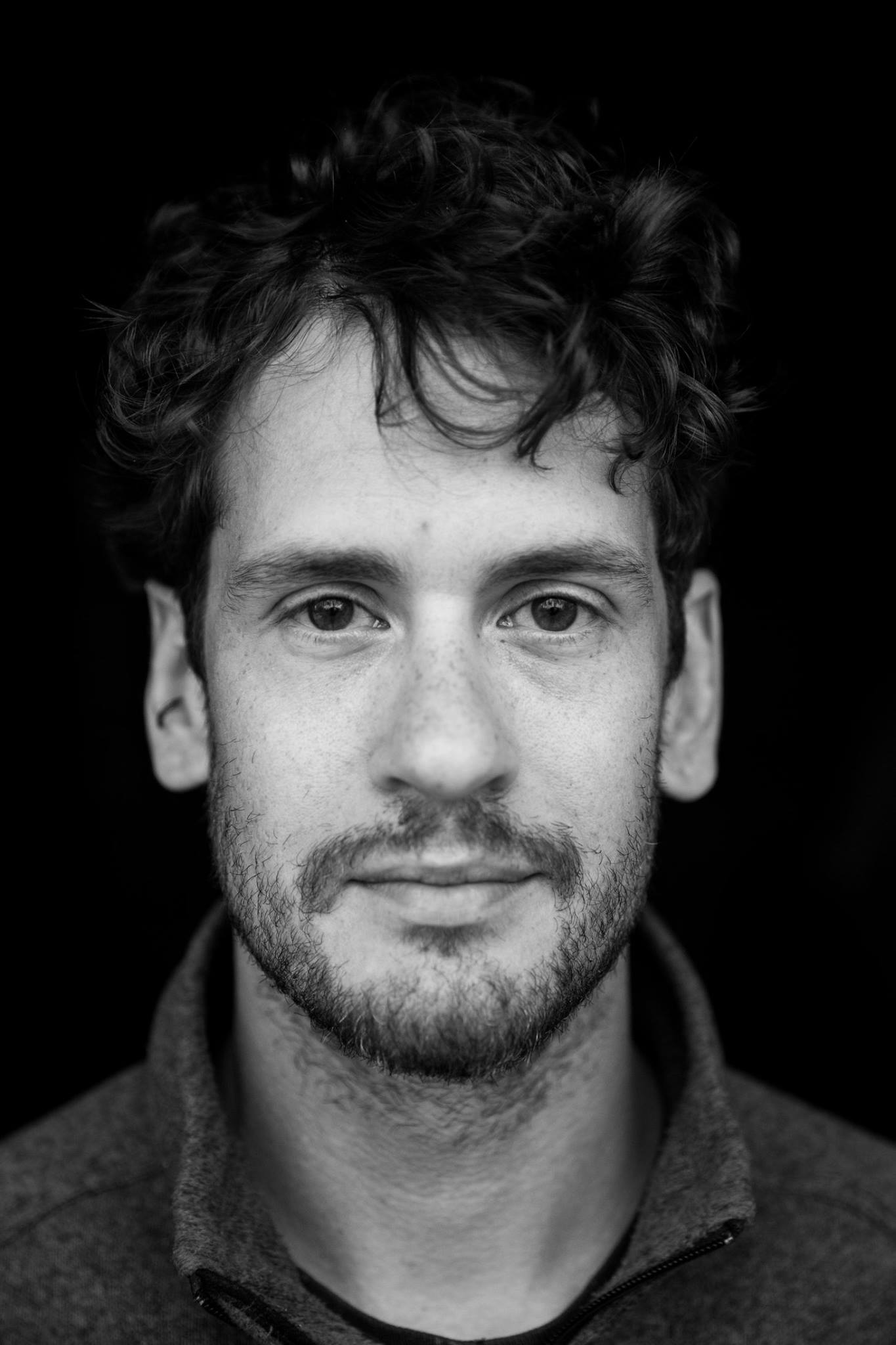}}]{Christopher~Robbiano} 
Christopher Robbiano received a BS in Physics and BS and MS degrees in Electrical Engineering at Colorado State University in 2011 and 2017, respectively.  He is currently a PhD candidate where his research has been in the areas of interactive sensing for sonar systems.  He currently works as a research scientist at Information System Technologies Incorporated, and previously worked as a digital design engineer at Broadcom.  His research interests include machine learning, autonomous systems, and statistical signal processing.
\end{IEEEbiography}
\vskip -2\baselineskip plus -1fil
\begin{IEEEbiography}[{\includegraphics[width=1in, height=1.25in, clip,keepaspectratio]{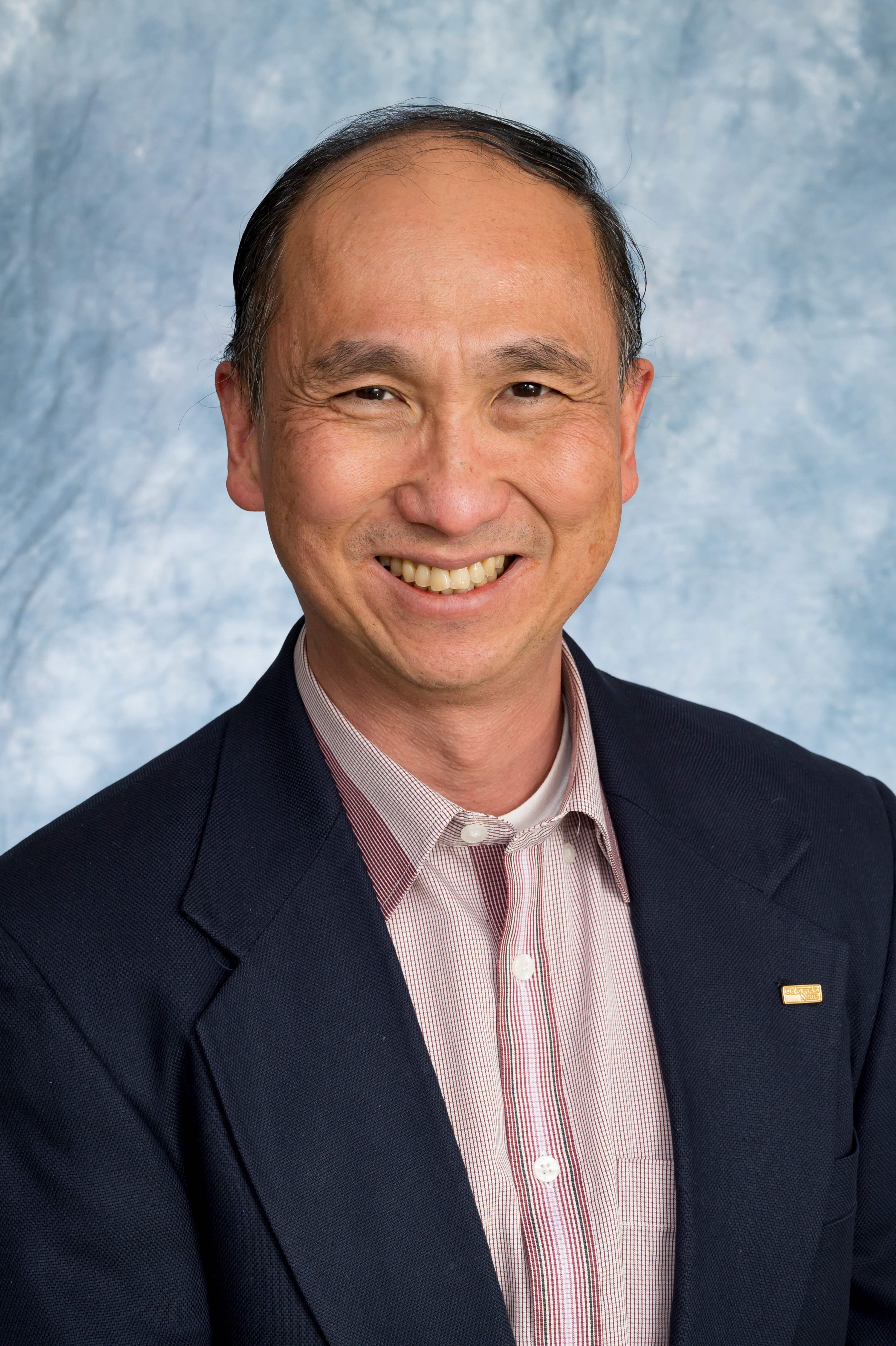}}]{Edwin~K.~P.~Chong}
Edwin K. P. Chong received his B.E.(First Class Honors) from the University of Adelaide, South Australia, in 1987; and his M.A. and Ph.D. in 1989 and 1991, respectively, both from Princeton, where he held an IBM Fellowship. He joined the School of Electrical and Computer Engineering at Purdue University in 1991, where he was named a University Faculty Scholar in 1999. Since August 2001, he has been a Professor of Electrical and Computer Engineering and of Mathematics at Colorado State University. He coauthored the best-selling book, An Introduction to Optimization (4th Edition, Wiley-Interscience, 2013). He is an IEEE Fellow and was President of the IEEE Control Systems Society in 2017.
\end{IEEEbiography}
\vskip -2\baselineskip plus -1fil
\begin{IEEEbiography}[{\includegraphics[width=1in, height=1.25in, clip,keepaspectratio]{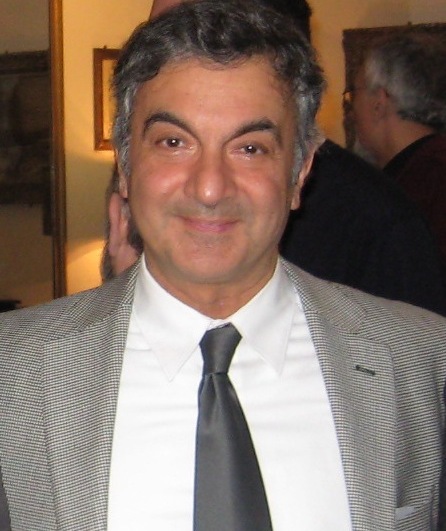}}]{Mahmood~R.~Azimi-Sadjadi}
Dr. Azimi-Sadjadi received his M.S. and Ph.D. degrees from the Imperial College of Science \& Technology, University of London, UK in 1978 and 1982, respectively, both in Electrical Engineering with specialization in Digital Signal/Image Processing. 

He is currently a full professor at the Electrical and Computer Engineering Department at Colorado State University (CSU). He is also serving as the director of the Digital Signal/Image Laboratory at CSU.   His main areas of interest include statistical signal and image processing, machine learning and adaptive systems, target detection, classification and tracking, sensor array processing, and distributed sensor networks.  

Dr. Azimi-Sadjadi served as an Associate Editor of the IEEE Transactions on Signal Processing and the IEEE Transactions on Neural Networks.   He is a Life Member of the IEEE. 
\end{IEEEbiography}
\vskip -2\baselineskip plus -1fil
\begin{IEEEbiography}[{\includegraphics[width=1in, height=1.25in, clip,keepaspectratio]{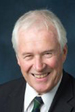}}]{Louis~L.~Scharf}
Louis Scharf is Research Professor of Mathematics and Emeritus Professor of Electrical and Computer Engineering at Colorado State University, Fort Collins, CO. His research interests are in statistical signal processing and machine learning, as it applies to adaptive array processing for radar, sonar, and communication; modal analysis for electric power monitoring; and image processing for classification. He has co-authored the books, L.L. Scharf, ``Statistical Signal Processing: Detection, Estimation, and Time Series Analysis,'' Addison-Wesley, 1991, and P.J. Schreier and L.L. Scharf, ``Statistical Signal Processing of Complex-Valued Data: The Theory of Improper and Noncircular Signals,'' Cambridge University Press, 2010. Professor Scharf has received several awards for his professional service and his contributions to statistical signal processing, including the Technical Achievement and Society Awards from the IEEE Signal Processing Society (SPS); the Donald W. Tufts Award for Underwater Acoustic Signal Processing, the Diamond Award from the University of Washington, and the 2016 IEEE Jack S. Kilby Medal for Signal Processing. He is a Life Fellow of IEEE.
\end{IEEEbiography}
\vskip -2\baselineskip plus -1fil
\begin{IEEEbiography}[{\includegraphics[width=1in, height=1.25in, clip,keepaspectratio]{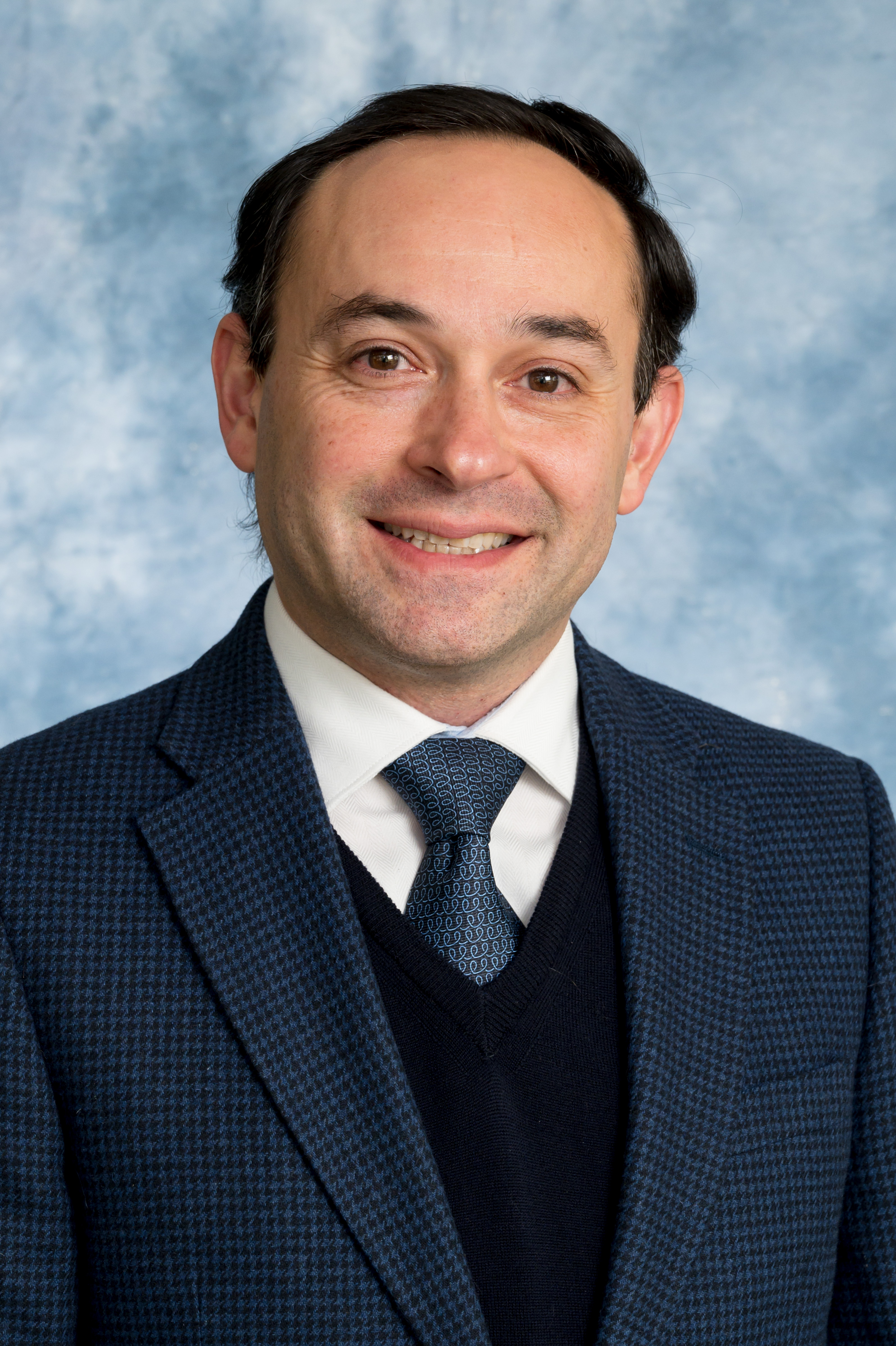}}]{Ali~Pezeshki}
Ali Pezeshki received the BSc and MSc degrees in electrical engineering from University of Tehran, Tehran, Iran, in 1999 and 2001, respectively. He earned his PhD degree in electrical engineering at Colorado State University in 2004. In 2005, he was a postdoctoral research associate with the Electrical and Computer Engineering Department at Colorado State University. From January 2006 to August 2008, he was a postdoctoral research associate with The Program in Applied and Computational Mathematics at Princeton University. In August 2008, he joined the faculty of Colorado State University, where he is now a Professor in the Department of Electrical and Computer Engineering, and the Department of Mathematics. His research interests are in statistical signal processing, machine learning, optimization, coding theory, geometry, applied harmonic analysis, and bioimaging. He served on the editorial board of IEEE ACCESS from 2012 to 2018.
\end{IEEEbiography}

\end{document}